\documentclass[11pt,superscriptaddress,floatfix]{article}
\pdfoutput=1
\usepackage{epsfig}
\usepackage{epstopdf}
\usepackage{graphicx,psfrag}
\usepackage{amssymb}

\usepackage[utf8]{inputenc}
\usepackage{jheppub}
\usepackage{color}
\usepackage{amsmath}
\usepackage{hyperref}
\usepackage{setspace}
\usepackage{array}
\usepackage{tikz}
\usepackage{amsthm}
\usepackage{bbold}
\usepackage[utf8]{inputenc}

\newtheorem{theorem}{Theorem}

\newcommand{\Tr}{\rm Tr}
\newcommand{\eref}[1]{(\ref{#1})}
\newcommand{\nn}{\nonumber}

\newcommand{\be}{\begin{eqnarray}}
\newcommand{\ee}{\end{eqnarray}}

\newcommand{\bmat}{\left ( \begin{array}{cc} }
\newcommand{\emat}{\end{array} \right ) }

\def\Tr{\textrm{Tr}}

 \def\firstcircle{(0,0) circle (1.5cm)}
  \def\secondcircle{(1.75,0) circle (1.5cm)}
  \def\thirdcircle{(0.85,1.75) circle (1.5cm)}
 \def\fourthcircle{(8.5,0) circle (1.5cm)}
  \def\fifthcircle{(10.25,0) circle (1.5cm)}
  \def\sixthcircle{(9.35,1.75) circle (1.5cm)}
   \def\seventhcircle{(0,0) circle (2cm)}
  \def\eighthcircle{(3,0) circle (2cm)}
  \def\ninthcircle{(1.5,2.5) circle (2cm)}

\newcommand{\beq}{\begin{equation}}
\newcommand{\beqs}{\begin{equation*}}
\newcommand{\eeq}{\end{equation}}
\newcommand{\eeqs}{\end{equation*}}

\begin{document}

\title{Exact moments of the Sachdev-Ye-Kitaev model up to order $1/N^2$}
\author[a]{Antonio M. Garc\'\i a-Garc\'\i a,}
\author[b]{Yiyang Jia}
\author[b]{and Jacobus J. M. Verbaarschot}
\affiliation[a]{Shanghai Center for Complex Physics, Department of Physics and Astronomy,
  Shanghai Jiao Tong University, Shanghai 200240, China}
\affiliation[b]{Department of Physics and Astronomy, Stony Brook University, Stony Brook, New York 11794, USA}
\emailAdd{amgg@sjtu.edu.cn}
\emailAdd{yiyang.jia@stonybrook.edu}
\emailAdd{jacobus.verbaarschot@stonybrook.edu}

\date{\today}
\abstract{
  We analytically evaluate the moments of the spectral density of the $q$-body Sachdev-Ye-Kitaev (SYK) model, and obtain order $1/N^2$  corrections for all moments,
  where $N$ is the total number of Majorana fermions.
  To order $1/N$, moments are given by those of the weight function of the Q-Hermite polynomials.
Representing Wick contractions by rooted
  chord diagrams, we show that the $1/N^2$ correction for each chord diagram
  is proportional to the number of triangular loops of the
  corresponding intersection
  graph, with an extra grading factor when $q$ is odd.
  Therefore the problem of finding $1/N^2$ corrections is mapped to a triangle counting problem. Since the  total number of triangles is a purely graph-theoretic property, we can compute
them for the  $q=1$ and $q=2$ SYK models, where the exact moments can be obtained analytically using other methods, and therefore we have solved the moment problem for any $q$ to
  $1/N^2$ accuracy. The moments are then used to obtain the spectral density of the SYK model to order $1/N^2$.
We also obtain an exact analytical result for all contraction diagrams contributing to the
moments, which can be evaluated up to eighth
  order. This shows that the Q-Hermite approximation is accurate even for small values of $N$.
  }
\maketitle\flushbottom
\newpage
\section{Introduction}

Although the study of strongly interacting quantum many body systems has a long history, many aspects still remain poorly understood. 
One of the difficulties is that the size of the Hilbert space
increases exponentially with the number of particles which severely limits
the scope of numerical studies. This is why analytical studies
of even simplified many-body systems contribute significantly to our
understanding of this problem. One such model is
the Sachdev-Ye-Kitaev (SYK) model \cite{sachdev1993,kitaev2015,maldacena2016}
which is a Hamiltonian system with an infinite range $q$-body
interaction acting on a many-body Hilbert space of Majorana fermions.
A similar model with
complex fermions was introduced several decades ago in the context
of nuclear physics where it became known as the two-body random ensemble
\cite{french1970,french1971,mon1975,bohigas1971,bohigas1971a,brody1981}. The motivation
for this model is that the nuclear interaction is mostly a two-body
interaction with matrix elements that appear close to random. It was also
known that  the level spacing distribution of nuclear levels
can be described by Random Matrix Theory \cite{wigner1951,dyson1962a,dyson1962b,dyson1962c,dyson1962d} , while the overall
shape of the nuclear level density does not resemble a semi-circle at all
but increases exponentially as $\exp( c \sqrt E)$ where $E$ is the energy
above the ground state \cite{bethe1936}. The two-body random ensemble addressed both of 
these issues, and has been studied intensively since then
 \cite{verbaarschot1984,mon1975,benet2001,benet2003,srednicki2002,small2014,small2015,borgonovi2016}.

 The recent interest in the SYK model \cite{sachdev1993,kitaev2015,maldacena2016,jensen2016,Jevicki:2016ito,Jevicki:2016bwu,Bagrets:2016cdf,cotler2016,Liu:2016rdi,Bagrets:2017pwq,Stanford:2017thb,altland2017,Witten:2016iux,Klebanov:2016xxf,Turiaci:2017zwd,Das:2017pif,Das:2017wae,krishnan2017} stems from the possibility that its gravity dual 
may be a quantum Anti-de Sitter space in two bulk dimensions (AdS$_2$) \cite{kitaev2015}. We note the possible relation between classical AdS$_2$ geometries and the SYK model was first proposed in Ref.\cite{sachdev2010}.
Both the SYK model and the AdS$_2$ gravity background are maximally chaotic \cite{kitaev2015,maldacena2016}, share
the same pattern of soft conformal symmetry breaking \cite{maldacena2016,maldacena2016a}, and similar low energy excitations \cite{maldacena2016,garcia2016,garcia2017} and low temperature thermodynamic properties \cite{parcollet1998,georges2001,maldacena2016,Jevicki:2016bwu,garcia2016}.
Since the SYK
model is analytically solvable for a large number of particles, including $1/N$ corrections \cite{maldacena2016},
this could provide us with a much deeper understanding of quantum aspects of the holographic duality beyond the usual large $N$ limit.

In previous works \cite{garcia2016,garcia2017}, two of us have studied both the
thermodynamic and spectral properties of the SYK model for $q>2$, and have
clearly established that the short-range spectral correlations are given by random
matrix theory which is a necessary ingredient for the model to be quantum chaotic
and therefore to have a gravity dual with black hole solutions. Moreover, it was found, by an explicit
analytical evaluation of the moments of the spectral density, that it grows
exponentially for low energies,
a typical feature of conformal field theories \cite{cardy1986} and therefore of gravity backgrounds with a field theory dual \cite{verlinde2000,carlip2005}.
One of the surprising results of these works is that the spectral density
at finite $N$, even for relatively small $N$, is very close to the weight function of the Q-Hermite polynomials. 

Rigorous results 
for the moments of a similar random spin
model \cite{erdos2014}, and very recently for the SYK model itself \cite{renjie2018}, show that in the large $N$ limit its spectral density
converges to the weight function of the Q-Hermite polynomials
only for $q \propto \sqrt N$ while in Refs.~\cite{garcia2016,garcia2017} $q$ 
was fixed and $N$ was relatively small, so such a good agreement was not expected.
Another surprising feature of the Q-Hermite approximation is that for low temperatures it
reproduces exactly the SYK partition function which in this limit reduces to the Schwarzian action and it is $1/N$ exact
\cite{maldacena2016,Bagrets:2016cdf,Bagrets:2017pwq,Stanford:2017thb}. This is again rather unexpected because this region 
is in principle controlled by high moments of order $N/q^2$ where deviations from the Q-Hermite result should be larger.  

  The main goal of the present paper is to study why the Q-Hermite approximation is so accurate.
  This question is addressed in two ways. First, by an analytical computation of the $1/N^2$ corrections to all moments
  which enables us to obtain the density up to that order for any $q$ , and second, by an exact analytical calculation of the finite-$N$ moments
  up to order eight.
  We note that originally the SYK model was only formulated for even $q$. However, it also makes sense for odd $q$, when the Gaussian-distributed operator becomes the supercharge of a supersymmetric
Hamiltonian \cite{fu2017,li2017,kanazawa2017,Peng:2017spg,Peng:2016mxj}. Unless stated otherwise, our results are valid for both
even and odd $q$, and for odd $q$ they refer to the spectral properties of
the supercharge.

We proceed by using the moment method for the spectral density.
The $1/N^2$ corrections are derived in two steps. First, we show that the $1/N^2 $ correction
to the Q-Hermite result for each contraction diagram is proportional to the total number of
triangular loops of the corresponding intersection graph. In the second step, we evaluate the
sum over all diagrams. This is a graph-theoretic problem with combinatorial factors that
can be determined from the exact expressions for the moments for $q=1$ and $q=2$.
The moments can be summed into a $1/N^2$ correction to the spectral density.

Finally, we note that other aspects of 
$1/N$ expansions in the SYK model were discussed in \cite{Ferrari:2017jgw,Benedetti:2017qxl,Dartois:2017xoe} but they
do not overlap with the present work.

This paper is organized as follows: In section \ref{sec:momMethod}, we define the SYK model, 
discuss the moment method and introduce the graphical representations
for the calculation of the moments.  In section \ref{sec:1/N} the $1/N$ expansion will be discussed, where
we also obtain the $1/N^2$ correction for a given diagram. 
In section \ref{sec:exact} we obtain a general and exact formula to evaluate the contraction diagrams.
In section \ref{sec:subsubcorrection}  we derive  triangle counting formulas, which in turn give us the total $1/N^2$ correction to moments.  After obtaining the $1/N^2$-exact moments, the corresponding
correction to the spectral density is evaluated in section \ref{sec:spectralDen}. In section \ref{sec:exactMom} we compute the exact sixth and eighth moments to further clarify the properties of the approximations we made. In section \ref{sec:qHermiteNature} we comment on the nature of the obtained results.
Concluding remarks and prospects for future work are discussed in
section \ref{sec:conclusions}.

\section{SYK model and moment method}\label{sec:momMethod}

\subsection{The SYK Hamiltonian}
 
The $q$-body SYK Hamiltonian is  given by
\begin{equation}
H(J_\alpha) = \sum_\alpha J_\alpha\Gamma_\alpha,
\end{equation}
where the   $\Gamma_\alpha$ are defined in terms of $2^{\lfloor N/2 \rfloor}\times 2^{\lfloor N/2 \rfloor}$ dimensional
Euclidean gamma matrices as
\begin{equation}
\Gamma_\alpha=(i)^{{q(q-1)}/{2}}\gamma_{i_1}\gamma_{i_2}\cdots\gamma_{i_q},
\end{equation}
with anti-commutation relations\footnote{We do not use Majorana convention $\{\gamma_k,\gamma_l\} = \delta_{kl}$ because we prefer $\gamma_k^2=1$. 
We can rescale to the Majorana convention by redefining the second moment, see equation (\ref{sigmaAndE0}).}
\be
\{\gamma_k,\gamma_l\} = 2\delta_{kl}.
\ee
The subscript $\alpha$ represents an index set with $q$ elements:
$\alpha=\{i_1,i_2,\ldots,i_q\}$, with $1\leq i_1<i_2<\cdots<i_q\leq N$. Hence $\alpha$ can have $\binom{N}{q}$ different configurations.
The couplings $J_\alpha$ are  random variables distributed according to
\begin{equation}\label{GaussianDistri}
P(J_\alpha)=\sqrt{\frac{N^{q-1}}{2(q-1)!\pi J^2}}\exp\left(-\frac{N^{q-1}J_\alpha^2}{2(q-1)!J^2}\right),
\end{equation}
where $J$ is a dimensionful parameter that sets the scale. Note we have included the factor $(i)^{q(q-1)/2} $ to  make the $\Gamma_\alpha$  Hermitian also for odd $q$, in which case $H(J_\alpha)$ is interpreted as the supercharge of the so called supersymmetric SYK model \cite{fu2017}.  

\subsection{Moments and Wick contractions}
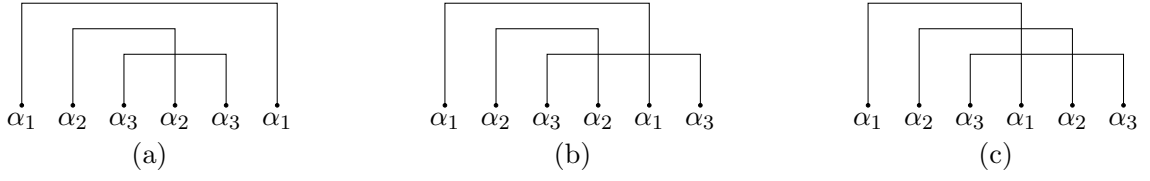
\begin{figure}[t!]
  \begin{center}
\begin{equation}
    \resizebox{\textwidth}{!}{
  \begin{tikzpicture}[scale=0.7]
\draw[fill=black] (0,0) circle (1pt);
\draw[fill=black] (1,0) circle (1pt);
\draw[fill=black] (2,0) circle (1pt);
\draw[fill=black] (3,0) circle (1pt);
\draw[fill=black] (4,0) circle (1pt);
\draw[fill=black] (5,0) circle (1pt);
\node at (0,-0.3) {$\alpha_1$};
\node at (1,-0.3) {$\alpha_2$};
\node at (2,-0.3) {$\alpha_3$};
\node at (3,-0.3) {$\alpha_2$};
\node at (4,-0.3) {$\alpha_3$};
\node at (5,-0.3) {$\alpha_1$};
\node at (2.5,-1) {(a)};

\draw[] (0,0) -- (0,2) -- (5,2) -- (5,0);
\draw[] (1,0) -- (1,1.5) -- (3,1.5) -- (3,0); 
\draw[] (2,0) -- (2,1) -- (4,1) -- (4,0);
\end{tikzpicture}\nn 
\hspace*{1.5cm}
\begin{tikzpicture}[scale=0.7]
\draw[fill=black] (0,0) circle (1pt);
\draw[fill=black] (1,0) circle (1pt);
\draw[fill=black] (2,0) circle (1pt);
\draw[fill=black] (3,0) circle (1pt);
\draw[fill=black] (4,0) circle (1pt);
\draw[fill=black] (5,0) circle (1pt);
\node at (0,-0.3) {$\alpha_1$};
\node at (1,-0.3) {$\alpha_2$};
\node at (2,-0.3) {$\alpha_3$};
\node at (3,-0.3) {$\alpha_2$};
\node at (4,-0.3) {$\alpha_1$};
\node at (5,-0.3) {$\alpha_3$};
\node at (2.5,-1) {(b)};
\draw[] (0,0) -- (0,2) -- (4,2) -- (4,0);
\draw[] (1,0) -- (1,1.5) -- (3,1.5) -- (3,0); 
\draw[] (2,0) -- (2,1) -- (5,1) -- (5,0);
\end{tikzpicture}\nn 
\hspace*{1.5cm}
\begin{tikzpicture}[scale=0.7]
\draw[fill=black] (0,0) circle (1pt);
\draw[fill=black] (1,0) circle (1pt);
\draw[fill=black] (2,0) circle (1pt);
\draw[fill=black] (3,0) circle (1pt);
\draw[fill=black] (4,0) circle (1pt);
\draw[fill=black] (5,0) circle (1pt);
\node at (0,-0.3) {$\alpha_1$};
\node at (1,-0.3) {$\alpha_2$};
\node at (2,-0.3) {$\alpha_3$};
\node at (3,-0.3) {$\alpha_1$};
\node at (4,-0.3) {$\alpha_2$};
\node at (5,-0.3) {$\alpha_3$};
\node at (2.5,-1) {(c)};

\draw[] (0,0) -- (0,2) -- (3,2) -- (3,0);
\draw[] (1,0) -- (1,1.5) -- (4,1.5) -- (4,0); 
\draw[] (2,0) -- (2,1) -- (5,1) -- (5,0);
\end{tikzpicture}
}
\end{equation}
\end{center}
\caption{Three contraction diagrams contributing to the
  sixth moment. } \label{fig1}
\end{figure}

An object of central interest is the spectral density $\rho(E)$:
\begin{equation}
\rho(E):=\left\langle \sum\limits_{k=1}^{2^{\lfloor \frac{N}{2} \rfloor}}\delta(E-E_{k}) \right \rangle,
\end{equation}
where $\langle \cdots \rangle$ denotes the ensemble average over the Gaussian distribution of $J_\alpha$. After a Fourier transform of the $\delta$-functions, we can write 
\begin{equation}
\begin{split}
\rho(E)&=\frac{1}{2\pi}\int_{-\infty}^{\infty}\text{d}te^{-iEt}\left\langle \text{Tr} e^{iHt}\right \rangle\\
&=\frac{1}{2\pi}\int_{-\infty}^{\infty}\text{d}te^{-iEt}\sum\limits_{k=0}^{\infty}\frac{(it)^k}{k!}\left\langle \text{Tr} H^k\right \rangle.
\end{split}
\end{equation}
Hence we can equivalently study the moments $\left \langle \text{Tr} H^k\right \rangle$. Due to the $J_\alpha\to -J_\alpha$ symmetry of the ensemble, all odd moments must vanish, and thus 
\begin{equation}
\rho(E)=\frac{1}{2\pi}\int_{-\infty}^{\infty}\text{d}te^{-iEt}\sum\limits_{p=0}^{\infty}\frac{(it)^{2p}}{(2p)!}\left\langle \text{Tr} H^{2p}\right \rangle.
\end{equation}
It will be convenient to factor out the dimensionality
of the Hilbert space and study the $2p$-th moment defined by
 \begin{equation}
  M_{2p}:= \left\langle \text{Tr} H^{2p}\right \rangle/2^{\lfloor N/2 \rfloor},
   \end{equation}
 and normalize the moments with respect to the second moment
 \be\label{eqn:scaledMoment}
 \frac { M_{2p}}{ M_2^p}.
 \ee
 It is easy to show 
 \be
 \label{eqn:m2}
 M_2=\binom{N}{q}\frac{(q-1)!J^2}{N^{q-1}} .
 \ee
 Since the average over the  $J_\alpha$'s is a Gaussian integration, the $2p$-th moment is given by the sume of all possible $(2p-1)!!$ Wick contractions among $p$ pairs of $\Gamma$'s. 
 A Wick contraction of the form, say,
 \be
 \label{eqn:GammaTrace}
 \text{Tr}(\Gamma_{\alpha_1}\Gamma_{\alpha_2}\Gamma_{\alpha_3}\Gamma_{\alpha_2}\Gamma_{\alpha_3}\Gamma_{\alpha_1}
 ),
\ee
where the Einstein summation convention is assumed, can be represented by diagram (a) in figure \ref{fig1}.
 We will call diagrams like figure \ref{fig1} contraction diagrams, which can be equivalently drawn as rooted chord diagrams on a circle \cite{flajolet2000}, and we will use the two terms interchangeably for such diagrams in this paper. 
 
The matrices $\Gamma_\alpha$'s satisfy
\begin{equation}\label{eqn:commrelation}
\Gamma_\alpha^2= 1, \quad \Gamma_\alpha\Gamma_\beta=(-1)^{q+c_{\alpha\beta}}\Gamma_\beta\Gamma_\alpha,
\end{equation} 
where there is no summation over repeated indices in the first equality,
and $c_{\alpha\beta}=|\alpha\cap\beta|$ is the number of common elements
between sets $\alpha$ and $\beta$.
We can use equation (\ref{eqn:commrelation}) to calculate traces of
products of $\Gamma_{\alpha_k}$ like  in (\ref{eqn:GammaTrace}) by
permuting $\Gamma_{\alpha_k}$'s until every two $\Gamma_{\alpha_k}$'s with the same subscript neighbor each other.
For intersecting neighboring contractions we thus have for fixed $\alpha$
\be
{N \choose q}^{-1}\sum_\beta \Gamma_\alpha \Gamma_\beta \Gamma_\alpha \Gamma_\beta =
{N \choose q}^{-1}\sum_{c_{\alpha\beta}=0}^q(-1)^{q+c_{\alpha\beta}} {N-q \choose q-c_{\alpha\beta}}{q \choose c_{\alpha\beta}} \mathbb{1}.
\label{commute}
\ee
We thus see that commuting two operators gives rise to the suppression factor
\be
\eta :={N \choose q}^{-1}\sum_{k=0}^q(-1)^{q+k} {N-q \choose q-k}{q \choose k},
\label{etadef}
\ee
which will play an essential role in the calculations of  this paper.
Using these relations, the trace in the example (\ref{eqn:GammaTrace}) can be written as
\be
\text{Tr}(\Gamma_{\alpha_1}\Gamma_{\alpha_2}\Gamma_{\alpha_3}\Gamma_{\alpha_2}\Gamma_{\alpha_3}\Gamma_{\alpha_1})/2^{\lfloor N/2 \rfloor}\
&=&\sum\limits_{\alpha_1,\alpha_2,\alpha_3}(-1)^{q+c_{\alpha_2\alpha_3}}  
= {N\choose q}^3 \eta.
\ee
Generically, a contraction with $p$ contraction lines can be written as 
\begin{equation}\label{eqn:phaseSum}
 (-1)^{q n_c}\sum\limits_{\alpha_1,\ldots,\alpha_p}(-1)^{\sum_{k=1}^{n_c} c_{\alpha_{i_k}\alpha_{j_k}}},
\end{equation}
where $n_c$ is the number of crossings in the contraction diagram, $\alpha_{i_k},\alpha_{j_k}$ belong to $\{\alpha_1,\ldots,\alpha_p\}$ and they label the contraction lines that cross each other.

\subsection{Intersection graphs}
The chord diagrams that contribute to the $2V$-th moment all have $V$ contraction lines/chords.
An intersection graph for a chord diagram with $V$ chords is defined as follows:
\begin{itemize}
\item Represent each chord by a vertex.
\item Connect two vertices with an edge if and only if there is a crossing between the two chords that these two vertices represent.
\end{itemize}

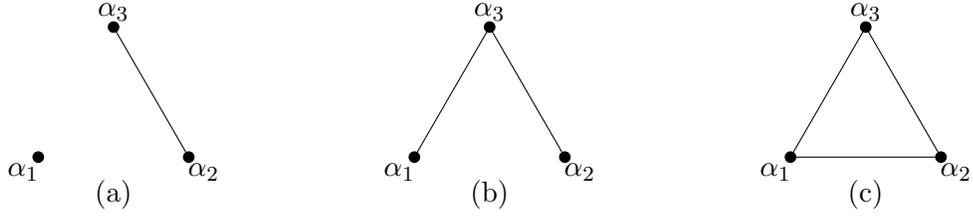
\begin{figure}
\begin{center}
  \begin{tikzpicture}
\draw[fill=black] (0,0) circle (2pt);
\draw[fill=black] (2,0) circle (2pt);
\draw[fill=black] (1,1.73) circle (2pt);

\draw[fill=black] (5,0) circle (2pt);
\draw[fill=black] (7,0) circle (2pt);
\draw[fill=black] (6,1.73) circle (2pt);

\draw[fill=black] (10,0) circle (2pt);
\draw[fill=black] (12,0) circle (2pt);
\draw[fill=black] (11,1.73) circle (2pt);
\node at (-0.2,-0.2) {$\alpha_1$};
\node at (2.2,-0.2) {$\alpha_2$};
\node at (1,1.93) {$\alpha_3$};
\node at (1,-0.5) {(a)};

\node at (4.8,-0.2) {$\alpha_1$};
\node at (7.2,-0.2) {$\alpha_2$};
\node at (6,1.93) {$\alpha_3$};
\node at (6,-0.5) {(b)};

\node at (9.8,-0.2) {$\alpha_1$};
\node at (12.2,-0.2) {$\alpha_2$};
\node at (11,1.93) {$\alpha_3$};
\node at (11,-0.5) {(c)};
\draw[] (2,0) -- (1,1.73);

\draw[] (7,0) -- (6,1.73) -- (5,0);
\draw[] (10,0) -- (11,1.73) -- (12,0) --(10,0);
  \end{tikzpicture}
\end{center}
  \caption{Intersection graphs corresponding to the contraction
    diagrams of figure \ref{fig1} in the same order from the left to
    the right. (a) has $V=3$ and $E=1$, (b) has $V=3$ and $E=2$, (c) has $V=3$ and $E=3$.} \label{fig2}
\end{figure}

We denote by $G$ a generic intersection graph, by $V$ the number of vertices and by $E$ the number of edges of an intersection graph. Therefore, in the notation of the previous section, $V=p$ and $E=n_c$. We give some examples of such diagrams in figure \ref{fig2}.

Motivated by the combinatorial factors that enter in the scaled moments
\eref{eqn:scaledMoment}, we define the following object associated with each contraction diagram and hence with each intersection graph, contributing to  the scaled $2V$-th moment,
\be\label{eqn:etaGDefinition}
\eta_G := (-1)^{Eq}\binom{N}{q}^{-V}\sum\limits_{\alpha_1,\ldots,\alpha_V}(-1)^{c(G)},
\ee
where $c(G)=\sum_{k=1}^{E} c_{\alpha_{i_k}\alpha_{j_k}}$, the $\alpha_{i_k}\alpha_{j_k}$ are all the edges in $G$ and $c_{\alpha_{i_k}\alpha_{j_k}}=|\alpha_{i_k}\cap\alpha_{j_k}|$. It is clear that an intersection graph $G$ completely determines the value of $\eta_G$. With the above definitions, we have 
\be
\frac{M_{2p}}{M_2^p}= \sum_{i=1}^{(2p-1)!!}\eta_{G_i}
\ee
by Wick's theorem, where $G_i$'s are the intersection graphs with $p$ vertices.

Notice that in the language of intersection graphs, $\eta$ defined in eq. \eref{etadef} corresponds to a $V=2, E=1$ graph, that is, a single edge connecting two vertices.

\subsection{Q-Hermite approximation}

Generally, a contraction diagram cannot be reduced to an expression only involving $\eta$ as we did in
\eref{commute}. The reason is that indices in more complicated contraction patterns
cannot be treated as being independent. However, we obtain an important approximation
if we nevertheless treat all crossings as independent: if a diagram has $E$ crossings the result is then simply given by \cite{garcia2017}
\be
\eta_G \approx \eta^E.
\label{qhapp}
\ee
 This approximation  expresses $\eta_G$ of an intersection graph $G$ by a product of its edges.
This approximation is in fact at least $1/N$-exact, as will be discussed in detail in section \ref{sec:1/N} and appendix \ref{proof:master}. It is exact for chord diagrams where multiple
crossings are indeed independent (having tree graphs as intersection graphs, see appendix \ref{append:CutVertex}).
The approximation (\ref{qhapp}) allows us to use  the Riordan-Touchard formula
\cite{riordan1975,touchard1952} to sum over all intersection graphs
\cite{garcia2016,cotler2016}:
\be\label{eqn:M2pRT}
\frac{M_{2p}}{M_2^p}=\sum_{i=1}^{(2p-1)!!}\eta_{G_i}\approx\sum_{i=1}^{(2p-1)!!}\eta^{E_i} = 
\frac{1}{(1-\eta)^p}\sum\limits_{k=-p}^{p}(-1)^k\eta^{k(k-1)/2}\binom{2p}{p+k},
\ee
where $E_i$ denotes the number of edges of $G_i$.
The unique spectral density that gives the moments  (\ref{eqn:M2pRT}) is the weight function of Q-Hermite polynomials \cite{ismail1987}:
 \begin{equation} \label{eqn:rhoQH}
\rho_{QH}(E) = c_N \sqrt{1-(E/E_0)^2}\prod\limits_{k=1}^{\infty}\left[1-4\frac{E^2}{E_0^2}\left(\frac{1}{2+\eta^k+\eta^{-k}}\right)\right],
\end{equation}
 with $c_N$ a normalization constant and $E_0$ a scale factor that drops out of the ratio
 \eref{eqn:scaledMoment}. 
 For this reason we refer to this approximation as the Q-Hermite approximation and also introduce the
 Q-Hermite moments
 \be
 \frac{M_{2p}^{\rm QH}}{M_2^p}:=\frac{1}{(1-\eta)^p}\sum\limits_{k=-p}^{p}(-1)^k\eta^{k(k-1)/2}\binom{2p}{p+k}.
 \ee
 
 We take equations (\ref{qhapp}), \eref{eqn:M2pRT} and \eref{eqn:rhoQH} as the first approximation to $\eta_G$,
   the moments and the spectral density respectively,
   and we use this as the starting point to investigate further $1/N^2$ corrections.
   We stress again that Q-Hermite approximation already contains many higher order terms in $1/N$, although the approximation is only exact to order $1/N$.

\section{$1/N$ expansion} \label{sec:1/N}
The goal of this paper is to understand better why the Q-Hermite result
discussed in the previous section is such a good approximation to the spectral density of the SYK model. This section is a step in this direction as we show that indeed there are
no $1/N$ corrections to the Q-Hermite moments, and give an argument that the $1/N^2$ corrections
are determined by the total number of triangles in an intersection graph.
In appendix \ref{proof:master}, we rigorously demonstrate this statement. 

The scaled Q-Hermite moments $M^{\text{QH}}_{2p}/M_2^p$ only depend on $\eta$ which
has the $1/N$ expansion
\be
\eta =(-1)^q\left( 1 -\frac {2q^2}N + \frac{2q^2(q-1)^2}{N^2}\right) +O\left(\frac{1}{N^3}\right).
\ee
Keeping only the leading power in $q$ at each order of $1/N$, it can be shown that this simplifies to
(see appendix~\ref{append:scalingLimitEta})
\be
\eta = (-1)^q\sum_{k=0}^\infty\frac 1{k!} \left( \frac {-2q^2}N \right )^k
= (-1)^qe^{-2q^2/N}.
\ee    
The Q-Hermite moments thus have a nontrivial large $N$ limit
when $q^2/N$ is kept fixed.
For $q \gg \sqrt N$ we have  $\eta \to 0$
so that only the nested contractions contribute, which give the moments
of a semi-circle. For $ q \ll \sqrt N $ we have to distinguish even and
odd $q$. For even $q$
we have $\eta \to 1$  so that
all contractions contribute equally which gives
the moments of a Gaussian distribution while in the case of odd $q$ we
obtain $\eta \to -1$ corresponding to the moments of the sum of two
delta functions located symmetrically about zero. In the latter case,
all scaled moments are equal to one (see  eq. \eref{phipmin1}).

To understand the corrections
to the Q-Hermite moments we evaluate $1/N$ corrections at fixed $q$ and $p$,
\be
\frac{M_{2p} - M^{\rm QH}_{2p}}{M_2^p} = \frac 1N a_1(p,q) + \frac 1{N^2} a_2(p,q) +\cdots . 
\ee
The Q-Hermite result is obtained when all crossings between contraction lines are
treated independently with each crossing contributing a factor $\eta$.
Corrections of order  $1/N$  occur when two crossed contraction lines have 
one common index.
Since this involves a single crossing, this correction is the same for the exact
result and the Q-Hermite result and we thus have that
\be
a_1(p,q) = 0.
\ee

Corrections to the Q-Hermite result occur when the crossed contracting
lines can no longer be permuted independently. Generally, this happens
when intersection graphs have closed loops,
and all vertices in the closed loop have at least pairwise common indices.
If a pair of vertices
does not have any common indices, they can be commuted or anti-commuted resulting in a loop
that is no longer closed and is thus given by the Q-Hermite results.
A closed loop of length $k$, thus differs by $O(1/N^{k-1})$ from the Q-Hermite
result. Therefore, for the $O(1/N^2)$ correction
we only need to consider the 
 triangular closed loops. For a closed loop of
three crossed contraction lines, say $\alpha$, $\beta$ and $ \gamma$, let
us consider the crossed pair $\beta\gamma$ with one common index
and let the crossed pair $\alpha\beta$ also have a
common index. This is a $1/N$ correction that is part of the Q-Hermite
result, and thus contributes as $q^2/N$ to leading order in $1/N$.
Deviations from the Q-Hermite result to
this closed loop occur when also the crossed pair
$\alpha\gamma$ has a common index. Choosing this index of $\gamma$ to be one from
$\alpha$ gives a second factor $q/N$. The index can either be among the
indices shared with $\beta$ or not. We thus conclude that the $1/N^2$
corrections due to a triangular diagram occur as
\be
\eta_G-\eta^{E} \sim \frac {q^3}{N^2}. 
\label{qn2}
\ee
The proportionality factor in \eref{qn2} can be obtained from the
simplest triangular intersection graph, whose value is referred to as $T_6$, see figure \ref{fig1} (c). This graph first occurs in the calculation of the sixth moment and can be calculated by keeping
track of the combinatorial factors \cite{garcia2016}
(see section \ref{sec:exactMom} for more details).
From the large $N$ expansion of $T_6$ and $\eta$ we then find
\be
T_6 - \eta^3 = -(-1)^{q}\frac {8q^3}{N^2} +O(1/N^3). 
\label{qn2ex} 
\ee
Therefore, this correction  is $1/\sqrt N$ suppressed in the large-$N$ limit with fixed $q^2/N $. Since for the lowest non vanishing order, 
 the triangles in the intersection graphs contribute independently,
 we arrive at the first main result of this paper:
\be
\eta_G-\eta^E = - \frac{8q^3}{N^2}(-1)^{qE}T +O(1/N^3),
\label{mom1n2}
\ee
where $T$ is the total number of triangles that occur in the
intersection graph $\eta_G$. This tells us the total  $1/N^2$ correction is obtained by counting the total number of triangles in all intersection graphs.
 In appendix \ref{proof:master} we will prove (\ref{mom1n2}) by a calculation
of the
$1/N^2$ corrections starting from  an exact formula for all Wick contractions which
will be derived in section \ref{sec:exact}.

In the double scaling limit, the Q-Hermite result for the moments only
depends on $q^2/N$. Therefore in this case the $1/N$ expansion is really in terms of this quantity only.
Corrections to a term of order $(q^2/N)^k$ occur when we fix additional
indices in a closed loop.
Choosing the remaining
indices gives a combinatorial factor of the form (with $m$ an integer satisfying $m \ll N$)
\be
   {N \choose q}^{-1} {N-m \choose q-1} \sim \frac qN.
   \ee

   For completeness we also give the first three terms of the $1/N$ expansion of the Q-Hermite approximation of $\eta_G$ which follows from the $1/N$ expansion of $\eta$. For a contraction diagram 
with $E$ crossings the we find
\be
\label{qh2}
\eta_G\approx\eta^{E} = (-1)^{E q} \left ( 1 - \frac {2Eq^2}N
+(2E^2q^4-4Eq^3+2Eq^2)\frac 1{N^2}\right ) +O(1/N^3).
\ee

\section{Exact result for the contraction diagrams}
\label{sec:exact}
In this section we derive an exact analytical expression for all contraction diagrams
contributing to the  moments of the SYK model.
We will use this result to prove \eref{mom1n2} by
an explicit calculation of the $1/N^2$ corrections, which we defer to appendix~\ref{proof:master} because this is a tedious calculation. The results of this section can also be used to obtain exact analytical results for low order
moments and some examples are worked out in appendix~\ref{append:q=2Moments}.

Since the phase factor $c(G)$  that appears in the definition of
$\eta_G$  (see eq. \eref{eqn:etaGDefinition}) is dependent on the number
of common elements in the index sets, it is natural to write
the combinatorics also in terms  of intersections of sets. Although $c(G)$ is determined by intersections of pairs, the combinatorics will
depend on intersections of arbitrary number of sets, so we introduce the objects
\be
c_{\alpha_1 \cdots \alpha_l}:=|\alpha_1 \cap \alpha_2 \cdots \cap \alpha_l|,
\ee
which are the number of  common indices in the vertices
$\{\alpha_1,\cdots,\alpha_l\}$, and
\be
d_{\alpha_1 \cdots \alpha_l},
\ee
which are the number of  common indices in the vertices
$\{\alpha_1,\cdots,\alpha_l\}$  that are not 
shared  with any of the other $\alpha_k$. By convention $\alpha_i$ and $\alpha_j$ cannot label the same vertex if $i\ne j$.
\begin{figure}[t!]
\begin{center}
\begin{tikzpicture}
      \begin{scope}
    \clip \secondcircle;
    \fill[red] \thirdcircle;
      \end{scope}
    \draw (-2.5,-2)--(-2.5,3.75)--(4.25,3.75)--(4.25,-2)--(-2.5,-2);
      \draw \firstcircle (-0.25,-0.25) node {$\alpha_1$};
      \draw \secondcircle (2.25,-0.25) node {$\alpha_2$};
      \draw \thirdcircle (0.85,2.25) node  {$\alpha_3$};
\draw (1,-2.5) node {Red region has cardinality $c_{\alpha_2\alpha_3}$};
 \begin{scope}
    \clip \fifthcircle;
    \fill[red] \sixthcircle;
      \end{scope}
      \begin{scope}
      \clip \fourthcircle;
    \clip \fifthcircle;
    \fill[white] \sixthcircle;
      \end{scope}
    \draw (6,-2)--(6,3.75)--(12.75,3.75)--(12.75,-2)--(6,-2);
      \draw \fourthcircle (8.25,-0.25) node {$\alpha_1$};
      \draw \fifthcircle (10.75,-0.25) node {$\alpha_2$};
      \draw \sixthcircle (9.35,2.25) node  {$\alpha_3$};
      \draw (9.5,-2.5) node {Red region has cardinality $d_{\alpha_2\alpha_3}$};
    \end{tikzpicture}
\end{center}
\caption{Venn diagrams with three index sets. Each index set is represented by a circle, containing $q$ elements. The box is the set of all possible values an index can take, which has cardinality $N$. The box is partitioned into eight regions.}\label{fig:indexVenn}
\end{figure}
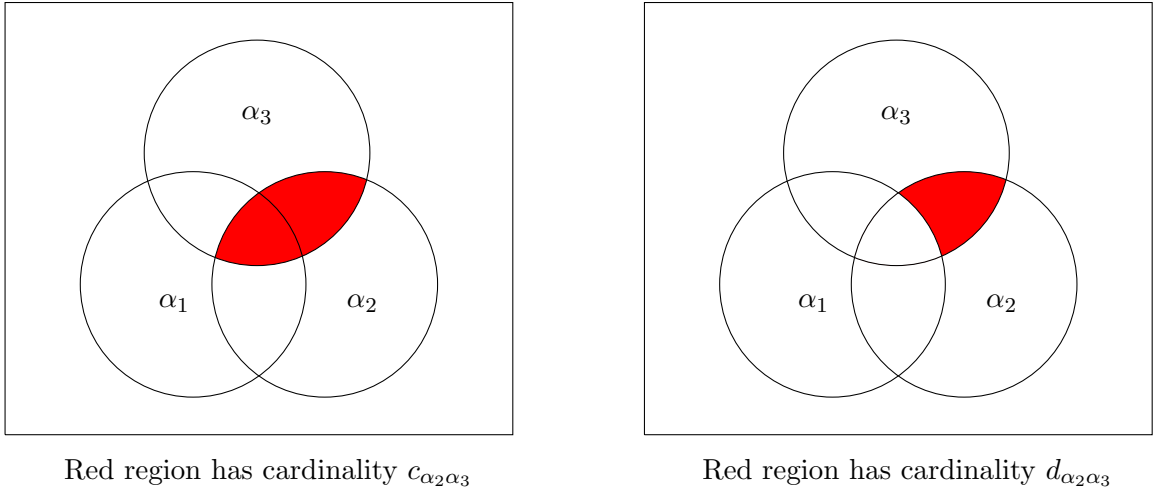

\begin{figure}[t!]
\begin{center}
\begin{tikzpicture}
      
    \draw [red] (-3.5,-2.5)--(-3.5,5.5)--(6.5,5.5)--(6.5,-2.5)--(-3.5,-2.5);
      \draw [red] \seventhcircle (-0.75,0.25) node {$\alpha_1$};
      \draw [red]\eighthcircle (3.25,0.25) node {$\alpha_2$};
      \draw [red] \ninthcircle (1.5,3.75) node  {$\alpha_3$};
      \draw (-0.3,-0.5) node {$q-d_{\alpha_1\alpha_2}-d_{\alpha_1\alpha_3}$};
      \draw (-0.3,-0.9) node {$-d_{\alpha_1\alpha_2\alpha_3}$};
      \draw (3.5,-0.5) node {$q-d_{\alpha_1\alpha_2}-d_{\alpha_2\alpha_3}$};
      \draw (3.5,-0.9) node {$-d_{\alpha_1\alpha_2\alpha_3}$};
            \draw (1.5,3) node {$q-d_{\alpha_1\alpha_3}-d_{\alpha_2\alpha_3}$};
      \draw (1.5,2.6) node {$-d_{\alpha_1\alpha_2\alpha_3}$};
      \draw (1.5,0.8) node {$d_{\alpha_1\alpha_2\alpha_3}$};
      \draw (2.5,1.5) node {$d_{\alpha_2\alpha_3}$};
      \draw (1.5,0) node {$d_{\alpha_1\alpha_2}$};
      \draw (0.5,1.5) node {$d_{\alpha_1\alpha_3}$};
      \draw (2,5) node {$N-[3q-(d_{\alpha_1\alpha_2}+d_{\alpha_2\alpha_3}+d_{\alpha_1\alpha_3})-2d_{\alpha_1\alpha_2\alpha_3}]$};
    \end{tikzpicture}
\end{center}
\caption{Venn diagrams with three index sets. There are eight regions, each labeled by its own cardinality. }\label{fig:indexVennPartitions}
\end{figure}
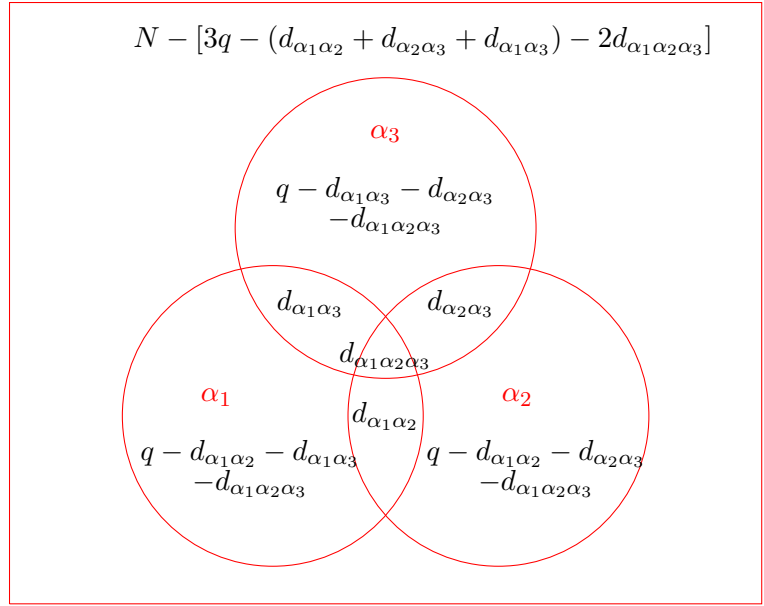

Essentially, the $c_{\alpha_1\cdots \alpha_k}$ and $d_{\alpha_1\cdots\alpha_k}$ are the
cardinalities of certain regions in the Venn diagram of $\{\alpha_1,\cdots,\alpha_l\}$. Figure~\ref{fig:indexVenn} illustrates the difference between $c_{\alpha_1\cdots \alpha_k}$ and $d_{\alpha_1\cdots \alpha_k}$ in the case of three index sets which
occur in the calculation of the sixth moment.
By the inclusion-exclusion principle, the two objects are
related by
\be\label{eqn:dcrelation}
d_{\alpha_1 \cdots \alpha_l}=c_{\alpha_1 \cdots \alpha_l}-c_{\alpha_1 \cdots \alpha_l*}+
c_{\alpha_1 \cdots \alpha_l**}-\cdots,
\ee
and conversely,
\be\label{eqn:cdrelation}
c_{\alpha_1 \cdots \alpha_l}=d_{\alpha_1 \cdots \alpha_l}
+d_{\alpha_1 \cdots \alpha_l*}+
d_{\alpha_1 \cdots \alpha_l**}+\cdots,
\ee
where stars in the subscripts indicate sums
over the remaining indices, e.g.
\be
c_{\alpha_1 \cdots \alpha_p**} = \sum_{\alpha_k\alpha_l\notin \{\alpha_1,\cdots,\alpha_p\}}c_{\alpha_1 \cdots \alpha_p\alpha_k\alpha_l},
\ee
and the same definition goes for the $d_{\alpha_1 \cdots \alpha_p*\cdots*}$. 
Eq. (\ref{eqn:cdrelation}) implies that $c(G)$ can be written in terms of $d$'s, hence we can write $\eta_G$ as 
\be
\binom{N}{q}^{V}(-1)^{qE}\eta_G= \sum\limits_{\{d_{\alpha_k\alpha_l}\}}
   \sum\limits_{\{ d_{\alpha_k\alpha_l\alpha_m}\}}
   \cdots\sum\limits_{\{d_{\alpha_1\ldots\alpha_{V}}\}}(-1)^{c(G)} {\cal M}
   .
\ee
The multiplicity factor ${\cal M}$ is the number of configurations that the index sets $\{\alpha_1,\ldots,\alpha_{V}\}$ can take given the values of the
$d_{\alpha_1\cdots\alpha_k} $.
  In general, a Venn diagram of $V$ index sets is partitioned into $2^{V}$ regions by the boundaries of the index sets, hence the multiplicity factor is the number of ways to distribute $N$ elements into $2^{V}$ regions, each region with its own cardinality. If the cardinality of each region is given by $m_i$, then
the multiplicity is given by the multinomial factor
\be
   {\cal M}= \frac{N!}{\prod_{i=1}^{2^{V} }m_i!}.
\ee
As an example,  figure~\ref{fig:indexVennPartitions} explicitly shows the cardinalities of all eight regions partitioned by the boundaries of three index sets.

Our final result for
the contribution of a given contraction diagram is thus given by,
\begin{align}
\label{eqn:etaGdCombinatorics}
(-1)^{qE}\eta_G=& \binom{N}{q}^{-V}
   \sum\limits_{\{d_{\alpha_k\alpha_l}\}}
   \sum\limits_{\{ d_{\alpha_k\alpha_l\alpha_m}\}}
   \cdots\sum\limits_{\{d_{\alpha_1\ldots\alpha_{V}}\}}
   (-1)^{c(G)} \frac{N!}{(N-Vq+d_2+2d_3+3d_4+\cdots)!}\nn \\
   & \times\prod_{k=1}^{V}\frac{1}{(q-d_{\alpha_k*}-d_{\alpha_k**}-\cdots)!}
   \prod_{1\leq i<j\leq V}\frac{1}{d_{\alpha_i\alpha_j}!}
   \prod_{1\leq i<j<k\leq V}\frac{1}{d_{\alpha_i\alpha_j\alpha_k}!}
   \cdots \frac{1}{d_{\alpha_1\alpha_2\cdots\alpha_{V}}!},
\end{align}
where
\be
\begin{split}
&d_2:=\sum_{1\le i<j\le V}d_{\alpha_i\alpha_j},\\
&d_3:=\sum_{1\le i<j<k\le V}d_{\alpha_i\alpha_j\alpha_k}
\end{split}
\ee
and so on. The expression
\be
Vq-d_2-2d_3-3d_4-\cdots
\ee
that appears in the denominator of the first factor for the multiplicity, is the cardinality of the union of all index sets, i.e. of all the circles in a Venn diagram like figure \ref{fig:indexVennPartitions}. One way to see this is
by noticing that the indices of the sets with cardinality $d_k$ are shared by $k$ of the $\alpha_k$ and that $k-1$ of them are not new.
The expression \eref{eqn:etaGdCombinatorics} is the second main result of this
paper.

The general expression \eref{eqn:etaGdCombinatorics} is a sum over $2^V -V-1$ variables, which 
limits its practical applicability. However, it can be simplified in several cases. First, nested
contractions, which correspond to isolated vertices in an intersection graph, just contribute a multiplicative factor of 1. So we do not need to include the sums over
isolated vertices in \eref{eqn:etaGdCombinatorics}, and this reduces the number of sums for these diagrams. 
A second simplification
 occurs if parts of an intersection graph are only
connected by a single vertex (see appendix \ref{append:CutVertex}). In that case the graph factorizes and each part can be evaluated separately by the formula \eref{eqn:etaGdCombinatorics}.

The $1/N$ corrections are controlled by the term 
\be \label{eqn:1/Ncontroll}
{N\choose q}^{-V}\frac{N!}{(N-Vq+d_2+2d_3+\cdots)!} \sim N^{-d_2-2d_3-\cdots}.
\ee
in \eref{eqn:etaGdCombinatorics}. Hence we have arrived at a convenient starting point for large $N$ expansions.

\section{$1/N^2$ corrections to the moments}\label{sec:subsubcorrection}

In section \ref{sec:1/N} we have seen that the $1/N^2$ corrections for
each graph are given by the number of triangles in an intersection
graph. To obtain the $1/N^2$ corrections to the moments, we have to
find the total number triangles in all intersection graphs contributing
to the moment of a given order. If $T_i$ is the number of triangles
in an intersection graph $G_i$  with $E_i$ edges, we have to evaluate
\be
\frac{M_{2p}-M_{2p}^{\text{QH}}}{M_2^p} =
-\frac {8q^3}{N^2} \sum_{i=1}^{(2p-1)!!} (-1)^{q{E_i}}{T_i}+O\left(\frac{1}{N^3}\right).
\label{stri}
\ee

In table \ref{tab1} we give the numerical results up to $2p =18$.
\begin{table}
\begin{center}
\begin{tabular}{|c|c|c|c|c|c|c|c|c|c|}
\hline $p$ & 1 & 2 & 3 & 4 & 5 & 6 & 7 & 8 & 9\\ \hline
$\sum_{i}T_i$& 0 & 0 & 1 & 28 & 630 & 13680 & 315315 & 7567560 & 192972780\\ \hline
$\sum_{i}(-1)^{E_i}T_i$ & 0 & 0 & -1 & -4& -10 & -20 & -35 & -56 & -84\\ \hline
\end{tabular}
\caption{The sum \eref{stri} for even $q$ and odd $q$ up to $2p =18$.}
\end{center}
\label{tab1}
\end{table}
Both for even $q$ and odd $q$ strikingly simple patterns emerge:
\begin{align}
 \sum\limits_{i=1}^{(2p-1)!!}T_i &= \frac{1}{15}\binom{p}{3}(2p-1)!!, \label{qEvenConjecture}\\
\sum\limits_{i=1}^{(2p-1)!!} (-1)^{E_i}T_i &= - \binom{p}{3},
  \label{qOddConjecture}
\end{align}
where $T_i$ and $E_i$ are the numbers of triangles and edges of the $i$-th intersection graph $G_i$.
These identities, which are one of the main results of this paper, will be
proved in the second part of this section.
 To order $1/N^2$, the moments
are thus given by
\begin{equation}\label{momqeven}
\frac{M_{2p}}{M_2^p}=\frac{1}{(1-\eta)^p}\sum\limits_{k=-p}^{p}(-1)^k\eta^{k(k-1)/2}\binom{2p}{p+k}-(2p-1)!!\binom{p}{3}\left(\frac{8q^3}{15N^2}\right)
\end{equation}
for even $q$
and by
\begin{equation}
  \label{supersymmMomentsConjecEqn}
\frac{M_{2p}}{M_2^p}=\frac{1}{(1-\eta)^p}\sum\limits_{k=-p}^{p}(-1)^k\eta^{k(k-1)/2}\binom{2p}{p+k}+\binom{p}{3}\left(\frac{8q^3}{N^2}\right)
\end{equation}
for odd $q$. 

The proof of (\ref{qEvenConjecture}) and (\ref{qOddConjecture}) is based on the following simple idea:
on the one hand, the  counting of the number of triangles occurring in intersection graphs
contributing to the $2p$-th moment is a graph-theoretic quantity, which is independent of the SYK model parameter $q$
(except for the parity of $q$); on the other hand, the SYK model for $q=1$ and $q=2$
is exactly solvable, which means that for $q=1$ and $q=2$ we can obtain $(M_{2p}-M_{2p}^{\text{QH}})/M_2^p$ to order $1/N^2$ from the exact solutions. Then
the proof follows
by  matching both sides of equation \eref{stri}.

We first consider the simpler case $q=1$, where
all moments are known analytically \cite{kieburg}. Since for $q=1$
\be
H=\sum_{\alpha=1}^NJ_\alpha\gamma_\alpha,
\ee
we have 
\be
H^2 = \sum_{\alpha=1}^N J_\alpha^2 \mathbb{1} , 
\ee
and 
\be
H^{2p}= \left(\sum_{\alpha=1}^N J_\alpha^2\right)^{p}\mathbb{1}.
\ee
Since $J_\alpha$ is Gaussian distributed, $\langle\Tr(H^{2p})\rangle$ can be easily calculated. In fact we can easily recognize it as the $p$-th moment of $\chi^2$ distribution with $N$ degrees of freedom
and the result is standard:
\be
\frac{M_{2p}^{q=1}}{ M_2^p} = \frac{\Gamma\left(\frac N2 +p\right)}{ \left(\frac N2\right)^p\Gamma\left(\frac N2\right)}
= 1+\frac{p(p-1)}N +  \left ( \frac{p^4}{2}
  -\frac{5p^3}{3}+\frac{3p^2}{2}-\frac{p}{3}\right ) \frac 1{N^2}+O(1/N^3).
\label{momq=1}
\ee
For  $q=1$, the $1/N$ expansion of the Q-Hermite moments simplifies to
\be
\frac{M^{\text{QH}, q=1}_{2p}}{M_2^p} = \sum_{i} \eta^{E_i} =
\sum_{i} (-1)^{E_i}
\left(1-\frac{2E_i}{N}
+(2E_i^2-2E_i)\frac{1}{N^2}\right)+O\left(\frac{1}{N^3}\right).
\ee
The total $1/N^2$ term is thus given by
\be
\frac{2}{N^2}\sum_{i}(-1)^{E_i}E_i(E_i-1)=\frac 2{N^2}\left. \frac {d^2 }{d\eta^2} \sum_{i} \eta^{E_i}\right |_{\eta=-1} .
\ee
The second derivative can be calculated analytically (see appendix \ref{append:edgeCounting}) and is given by
\be
\left. 2\frac {d^2 }{d\eta^2} \sum_{i} \eta^{E_i}\right |_{\eta=-1} \frac 1{N^2}
=12 {p \choose 4} \frac 1{N^2}.
\ee
Subtracting this from the exact $q=1$ result, eq. \eref{momq=1}  we find
\be
\frac{M_{2p}^{q=1} - M_{2p}^{{\rm QH}, q=1}}{M_2^p} = 8 {p \choose 3}\frac 1{N^2} = -\frac{8\times 1^3}{N^2}\sum_{i=1}^{(2p-1)!!} (-1)^{{E_i}}{T_i} .
\ee
This proves (\ref{qOddConjecture}).

For $q=2$ there is no compact formula for the $2p$-th moment, however we can still compute the moments to $1/N^2$ from the exact joint probability distribution of the coupling matrices. The computation is more involved and we give the derivation in appendix \ref{append:q=2Moments}, here we only quote the final result:
\be
\frac{M_{2p}^{q=2}}{M_2^p}=(2p-1)!!\left[1-\frac{8}{3}\binom{p}{2}\frac{1}{N}+\frac{8}{9}\binom{p}{2}(2p^2-2p-1)\frac{1}{N^2}\right]+O\left(\frac{1}{N^3}\right).
\ee
Meanwhile the Q-Hermite result is given by
 \be
\frac{M^{\text{QH}}_{2p}}{M_2^p} = \sum_{i} \eta^{E_i} =\sum_{i}  \left ( 1 - \frac {8E_i}N
+(32E_i^2-24E_i)\frac 1{N^2}\right ) +O(1/N^3).
\ee
Again the $1/N^2$ sum can be computed using the techniques explained in appendix \ref{append:edgeCounting}:
\be
\frac{1}{N^2}\sum_{i}(32E_i^2-24E_i)&=& \frac{1}{N^2}\left( \left. 32\frac {d^2 }{d\eta^2} \sum_{i} \eta^{E_i}\right  |_{\eta=1}+\left. 8\frac {d}{d\eta} \sum_{i} \eta^{E_i}\right  |_{\eta=1}\right)
\nn \\
&=&\frac{(2p-1)!!}{N^2}\left(\frac{8 p^4}{9}-\frac{16 p^3}{15}-\frac{76 p^2}{45}+\frac{28 p}{15}\right).
\ee
We finally find
\be
\frac{M_{2p}^{q=2} - M_{2p}^{{\rm QH}, q=2}}{M_2^p} = -\frac{64}{15} (2p-1)!!{p \choose 3}\frac 1{N^2}=  -\frac {8\times 2^3}{N^2}\sum\limits_{i=1}^{(2p-1)!!}T_i.
\ee
This proves (\ref{qEvenConjecture}).
\section{Corrections to the  spectral density}\label{sec:spectralDen}
The moments, both for even $q$ and odd $q$, satisfy Carleman's condition and hence
uniquely determine the spectral density \cite{Berg2002}. In the following two
subsections we give the spectral density
corrections for the even $q$ and the odd $q$ cases.

\subsection{Spectral density for even $q$}\label{sec:evenqSpecDen}
We decompose the spectral density into the weight function $\rho_{QH}(E)$
of the Q-Hermite polynomials,
determined by the Q-Hermite moments, plus a correction $\delta \rho(E)$ determined
by the $1/N^2 $ correction to the moments,
\begin{equation}
\rho(E)= \rho_{QH}(E)+\delta\rho(E).
\end{equation}
where  $\rho_{QH}(E)$  is given by \cite{ismail1987,garcia2017}
\begin{equation} \label{totalrhowithcorrection}
\rho_{QH}(E) = c_N \sqrt{1-(E/E_0)^2}\prod\limits_{k=1}^{\infty}\left[1-4\frac{E^2}{E_0^2}\left(\frac{1}{2+\eta^k+\eta^{-k}}\right)\right],
\end{equation}
and for even $q$  we have\footnote{The $2^q$ factor in $\sigma^2$ is to rescale the $\gamma$ matrices to the Majorana convention $\{\gamma_i,\gamma_j\}=\delta_{ij}$.} 
\begin{equation}\label{sigmaAndE0}
\begin{split}
&\sigma^2 = M_2=\binom{N}{q} \frac{J^2(q-1)!}{2^qN^{q-1}}\sim \frac{2^{-q}J^2}{q}N, \\
&E_0= \left(\frac{4\sigma^2}{1-\eta}\right)^{1/2}\sim \sqrt{\frac{2^{1-q}}{q^3}}JN.
\end{split}
\end{equation}
We  normalize $\rho_{QH}(E)$ by
\begin{equation}
\int \rho_{QH}(E)dE=2^{N/2}.
\end{equation}
This results in the 
normalization constant \cite{ismail1987}
\be
c_N = \frac {2^{N/2}}{\pi\sigma} (1+\eta)\sqrt{1-\eta}\prod_{k=1}^\infty \frac{1-\eta^{2k+2}}{1-\eta^{2k+1}}.
\label{cn}
\ee    
 After performing a Poisson resummation and ignoring certain exponentially small (in $N$) terms \cite{garcia2017}, the spectral density away from $|E|=|E_0|$ simplifies to 
\begin{equation}
\rho_{QH}(E) = c_N \exp\left[\frac{2\arcsin^2(E/E_0)}{\log\eta}\right]\left(1-\exp\left[-\frac{4\pi}{\log\eta}\left(|\arcsin(E/E_0)|-\frac{\pi}{2}\right)\right]\right).
\label{rhoqh}
\end{equation}
From this we deduce that at large $N$,
\begin{equation}
c_N\sim \frac{1}{\pi\sigma} 2^{N/2}.
\end{equation}
It is simple to verify that the  correction term 
\begin{equation}\label{rhocorrection}
\delta\rho(E)= -2^{N/2}\frac{q^3\sigma^5}{90\sqrt{2\pi}N^2}\frac{d^6}{dE^6}\exp\left({-\frac{E^2}{2\sigma^2}}\right)
\end{equation}
gives the moments \eref{momqeven} consistent with the normalization
of the $\rho_{\rm QH}(E)$.

For any fixed value of energy $E$, $E/E_0$ is small since $E_0\sim N$, and the leading behavior of $\rho_{QH}$ is given by 
\begin{equation}
\rho_{QH}(E)\sim \frac{1}{\sigma} 2^{N/2} \exp\left(-\frac{E^2}{2\sigma^2}\right),
\end{equation}
while the leading behavior of $\delta\rho$ is 
\begin{equation}\label{dengas}
  \delta\rho(E) \sim -\frac{1}{N^2 \sigma}2^{N/2}
  \exp\left(-\frac{E^2}{2\sigma^2}\right).
\end{equation}
This is indeed a small correction in the point-wise sense both for large $N$ at fixed $q$
and in the double scaling limit in terms of the variable $E/E_0\ll 1$.
Unfortunately, this is not a small correction in the uniform sense, for example, if instead of a fixed $E$ one looks at a fixed value of the scaling variable $x=E/E_0$, then for $x$ close to 1,
the  correction term
\be
\delta \rho(E) \sim 2^{N/2}e^{-N/q^2}
\ee
becomes exponentially larger than the leading term \eref{rhoqh}
\be
\rho_{QH}(E) \sim 2^{N/2}e^{-\pi^2 N/4 q^2}.
\ee
This prevents us from obtaining a meaningful correction to the free energy by integrating $\delta\rho$.
This is indeed consistent with the fact that the SYK partition function is $1/N$ exact in the low temperature limit \cite{maldacena2016,garcia2017,Stanford:2017thb,Bagrets:2017pwq} where it is 
dominated by the spectral density for $E \approx E_0$, so $1/N^2$ corrections in this region must be spurious.   

\subsection{Spectral density for odd $q$}

For odd $q$, $\eta < 0$, but the expression (\ref{totalrhowithcorrection})
for the Q-Hermite spectral density
is still applicable. Following the steps of the even $q$ calculation, it is 
straightforward to show that for 
large $N$ and away from the edge of the spectrum, the spectral density
\eref{rhoqh}
is given by \cite{Garcia-Garcia:2018ruf}
\be
\rho_{\rm QH} (E) = c_N  \cosh \frac {\pi\arcsin(E/E_0)}{\log|\eta|}\exp\left [ 2 
  \frac { \arcsin^2(E/E_0)}{\log|\eta|} \right ].
\label{rhoasym}
\ee
The normalization constant can be  determined from
\be
\int dE \rho_{\rm QH}(E) dE = 2^{N/2}.
\ee
In the large $N$ limit, the integral can be evaluated by a saddle point approximation. Using
that $\log| \eta| \sim  -2 q^2/N$ in this limit we find
\be
c_N = e^{N/2\log 2- N\pi^2/16 q^2},
\label{normodd}
\ee
which gives exactly the leading order $1/q^2$ correction to the zero temperature entropy
\cite{fu2017}.
In terms of units where the second moment is
normalized to one, the correction to the spectral density with moments given by
(\ref{supersymmMomentsConjecEqn}) is equal to
\be
\delta \rho(x) = \frac {2^{\frac N2}q^3}{N^2}\left[\frac{5}{2}\delta(x^2-1)+\frac{3}{2x}\frac{d}{dx}\delta(x^2-1)+\frac{1}{x^2}\frac{d^2}{dx^2}\delta(x^2-1)-\frac{1}{6x^3}\frac{d^3}{dx^3}\delta(x^2-1)\right].
  \ee
  In terms of physical units with $M_2=\sigma^2$ this can be written as
\be \label{dendelta}
\delta \rho(E) &=& \frac {2^{\frac N2}q^3}{N^2}\left[\frac{5}{2}\delta(E^2-\sigma^2)
  +\frac{3\sigma^2}{2E}\frac{d}{dE}\delta(E^2-\sigma^2)\right.+\frac{\sigma^4}{E^2}\frac{d^2}{dE^2}\delta(E^2-\sigma^2) \nonumber\\ 
&&\left.-\frac{\sigma^6}{6E^3}\frac{d^3}{dE^3}\delta(E^2-\sigma^2)\right].
\ee
So also for odd $q$ we find that the $1/N^2$ correction to the spectral density is
given by derivatives of its large $N$ limit.

  Correction terms in the form of $\delta$-functions are not strange to random
  matrix theory: for example the $1/N$ correction of the Wigner-Dyson
  ensemble is proportional to  $\delta$-functions at the edges of the
  semi-circle \cite{verbaarschot1984}.

 \section{Exact calculation of the sixth and eighth moment}\label{sec:exactMom}

 In this section we give exact results for the sixth and eighth moment. More details
 can be found in appendix \ref{append:gemnom}.
 
 The sixth moment was already calculated in \cite{garcia2016}. All diagrams
 except for the rightmost contraction diagram in figure \ref{fig1} coincide with the
 Q-Hermite result which allows the application of Riordan-Touchard formula
 \be
\sum_{i=1}^{(2p-1)!!} \eta^{E_i} =
 \frac 1{(1-\eta)^p} \sum_{k=-p}^p (-1)^k \eta^{k(k-1)/2}
       {2p \choose p+k}.
       \ee
       For the sixth moment the Q-Hermite result is given by 
       \be
      \frac{M_6^{\text{QH}}}{M_2^3}=5 + 6\eta +3 \eta^2 +\eta^3,
       \ee
       while the exact sixth moment is given by
       \be
       \frac{M_6}{M_2^3}=5 + 6\eta +3 \eta^2 +T_6,
       \ee
       with
       \be
       T_6 ={N\choose q}^{-2}
  \sum_{k=0}^q \sum_{m=0}^{q}(-1)^{q-k-m}
      {N-2k\choose q-m} {2k \choose m}{N-q \choose k}
      {q\choose k}.
      \ee
In table \ref{tab:6thMomIntersecGraphs} we list all contributions to the sixth moment.
      
\begin{table}[t!]
  \begin{center}
\begin{tabular}{|c|c|c|c|c|}
\hline Intersection graph & \begin{tikzpicture} \draw[fill=black] (0,0) circle (1pt); \draw[fill=black] (0.4,0) circle (1pt);\draw[fill=black] (0.2,0.346) circle (1pt); \node at (0.2,0.35) {};\end{tikzpicture} & \begin{tikzpicture} \draw[fill=black] (0,0) circle (1pt); \draw[fill=black] (0.4,0) circle (1pt);\draw[fill=black] (0.2,0.346) circle (1pt);
\draw (0,0)--(0.4,0); \end{tikzpicture} & \begin{tikzpicture} \draw[fill=black] (0,0) circle (1pt); \draw[fill=black] (0.4,0) circle (1pt);\draw[fill=black] (0.2,0.346) circle (1pt);
\draw (0,0)--(0.4,0)--(0.2,0.346); \end{tikzpicture} & \begin{tikzpicture} \draw[fill=black] (0,0) circle (1pt); \draw[fill=black] (0.4,0) circle (1pt);\draw[fill=black] (0.2,0.346) circle (1pt);
\draw (0,0)--(0.4,0)--(0.2,0.346)--(0,0); \end{tikzpicture} \\ \hline
Value & 1 & $\eta$ & $\eta^2$ & $T_6$\\ \hline
Multiplicity & 5 & 6 & 3 & 1\\ \hline
\end{tabular}
\caption{All the intersection graphs for the sixth moment.}\label{tab:6thMomIntersecGraphs}
\end{center}
\end{table}
      Again using the Riordan-Touchard formula we find the Q-Hermite result
      for the eighth moment
      \be
      \frac{M_8^{\rm QH}}{M_2^4} =       14 + 28 \eta + 28 \eta^2 + 20 \eta^3 +10 \eta^4 + 4\eta^5 +\eta^6 .
      \ee
      The exact result for the 8th moment is given
      by
      \be
      \frac{M_8}{M_2^4} = 14 + 28 \eta + 28 \eta^2 + 12 \eta^3 + 8 T_6
      +2 T_{44} +8\eta T_6 +4 T_{66} + T_8.
      \ee
      It involves three new structures (see the intersection graphs in Table \ref{tab:8thMomIntersecGraphs}).
      They are still simple enough that they can be expressed as simple sums by inspection. However, they can also
      be derived starting from the general formula \eref{eqn:etaGdCombinatorics} and we give two
      examples in appendix \ref{append:genmomex}.
      The first structure corresponding to the square intersection
      graph  (see table \ref{tab:8thMomIntersecGraphs}) is equal to
      \be
      T_{44} = {N \choose  q}^{-3}\sum_{k=0}^q \sum_{r=0}^{q}  \sum_{s=0}^{q} (-1)^{r+s}
    {N-q \choose k}{q \choose k}
    { N-2k \choose q-r}
    { 2k \choose r} 
    { N-2k \choose q-s}{2k  \choose s}.
    \ee
\begin{table}
  \begin{center}
\begin{tabular}{|c|c|c|c|c|c|c|c|c|c|c|c|}
\hline Intersection graph & \begin{tikzpicture} \draw[fill=black] (0,0) circle (1pt); \draw[fill=black] (0.4,0) circle (1pt);\draw[fill=black] (0.4,0.4) circle (1pt);\draw[fill=black] (0,0.4) circle (1pt); \node at (0.2,0.44) {};\end{tikzpicture} & \begin{tikzpicture}\draw[fill=black] (0,0) circle (1pt); \draw[fill=black] (0.4,0) circle (1pt);\draw[fill=black] (0.4,0.4) circle (1pt);\draw[fill=black] (0,0.4) circle (1pt);
\draw (0,0)--(0.4,0); \end{tikzpicture} & \begin{tikzpicture} \draw[fill=black] (0,0) circle (1pt); \draw[fill=black] (0.4,0) circle (1pt);\draw[fill=black] (0.4,0.4) circle (1pt);\draw[fill=black] (0,0.4) circle (1pt);
\draw (0,0)--(0.4,0);\draw (0,0.4)--(0.4,0.4); \end{tikzpicture} & \begin{tikzpicture} \draw[fill=black] (0,0) circle (1pt); \draw[fill=black] (0.4,0) circle (1pt);\draw[fill=black] (0.4,0.4) circle (1pt);\draw[fill=black] (0,0.4) circle (1pt);
\draw (0,0)--(0.4,0)--(0.4,0.4); \end{tikzpicture} & \begin{tikzpicture} \draw[fill=black] (0,0) circle (1pt); \draw[fill=black] (0.4,0) circle (1pt);\draw[fill=black] (0.4,0.4) circle (1pt);\draw[fill=black] (0,0.4) circle (1pt);
\draw (0,0)--(0.4,0)--(0.4,0.4);\draw (0.4,0)--(0,0.4); \end{tikzpicture}& \begin{tikzpicture} \draw[fill=black] (0,0) circle (1pt); \draw[fill=black] (0.4,0) circle (1pt);\draw[fill=black] (0.4,0.4) circle (1pt);\draw[fill=black] (0,0.4) circle (1pt);
\draw (0,0)--(0.4,0)--(0.4,0.4)--(0,0.4); \end{tikzpicture}& \begin{tikzpicture} \draw[fill=black] (0,0) circle (1pt); \draw[fill=black] (0.4,0) circle (1pt);\draw[fill=black] (0.4,0.4) circle (1pt);\draw[fill=black] (0,0.4) circle (1pt);
\draw (0,0)--(0.4,0)--(0.4,0.4)--(0,0); \end{tikzpicture}& \begin{tikzpicture} \draw[fill=black] (0,0) circle (1pt); \draw[fill=black] (0.4,0) circle (1pt);\draw[fill=black] (0.4,0.4) circle (1pt);\draw[fill=black] (0,0.4) circle (1pt);
\draw (0,0)--(0.4,0)--(0.4,0.4)--(0,0)--(0,0.4); \end{tikzpicture}& \begin{tikzpicture} \draw[fill=black] (0,0) circle (1pt); \draw[fill=black] (0.4,0) circle (1pt);\draw[fill=black] (0.4,0.4) circle (1pt);\draw[fill=black] (0,0.4) circle (1pt);
\draw (0,0)--(0.4,0)--(0.4,0.4)--(0,0.4)--(0,0); \end{tikzpicture}& \begin{tikzpicture} \draw[fill=black] (0,0) circle (1pt); \draw[fill=black] (0.4,0) circle (1pt);\draw[fill=black] (0.4,0.4) circle (1pt);\draw[fill=black] (0,0.4) circle (1pt);
\draw (0,0)--(0.4,0)--(0.4,0.4)--(0,0.4)--(0,0)--(0.4,0.4); \end{tikzpicture}& \begin{tikzpicture} \draw[fill=black] (0,0) circle (1pt); \draw[fill=black] (0.4,0) circle (1pt);\draw[fill=black] (0.4,0.4) circle (1pt);\draw[fill=black] (0,0.4) circle (1pt);
\draw (0,0)--(0.4,0)--(0.4,0.4)--(0,0.4)--(0,0)--(0.4,0.4);\draw (0,0.4)--(0.4,0); \end{tikzpicture}\\ \hline
Value & 1 & $\eta$ & $\eta^2$ & $\eta^2$ & $\eta^3$ & $\eta^3$ &$T_6$ & $\eta T_6$ & $T_{44}$& $T_{66}$ & $T_8$\\ \hline
Multiplicity & 14 & 28 & 4 & 24 & 4 & 8 & 8& 8& 2 & 4 &1\\ \hline
\end{tabular}
\caption{All the intersection graphs for the eighth moment.}\label{tab:8thMomIntersecGraphs}
\end{center}
\end{table}
The second structure corresponding to the square intersection diagram with one diagonal
(see table \ref{tab:8thMomIntersecGraphs}) only differs by an additional phase factor
    \be
    T_{66}  =  {N \choose  q}^{-3}\sum_{k=0}^q
    \sum_{r=0}^{q} \sum_{s=0}^{q} (-1)^{k+r+s}
    {N-q \choose k}{q \choose k}
    { N-2k \choose q-r}
    { 2k \choose r} 
    { N-2k \choose q-s}{2k  \choose s}.
    \ee
    The most complicated diagram is the one with 6 crossings corresponding to the rightmost
   graph in table \ref{tab:8thMomIntersecGraphs}. It is given by
\be
 T_8 &=& {N \choose  q}^{-3}\sum_{k =0}^q  \sum_{r=0}^q \sum_{s=0}^{q+r}
\sum_{t=0}^{s}
(-1)^{r+k+s+t}
 {N-q \choose k}{q \choose k}
 { 2k\choose s}{s \choose t} 
{N-2k \choose q+r-s} 
 {q+r-s \choose q-r-t}
 {2r\choose r}.\nn \\
\label{T8-simplea}
 \ee
The results for the sixth and eighth moment have been
 simplified using the convolution property of
 binomial factors. For example, for $T_8$ we initially obtain the result in the form of an
 8-fold sum. The general result
 gives an 11-fold sum which can be reduced to this result as
 is worked out in detail in appendix
 \ref{append:gemnom} where we also list all contraction diagrams contributing to $M_8$.

 The results for the moments are also valid for odd $q$, even for $q=1$.
 We have checked that the above expressions simplify to the $ q=1$ result
 in eq.~\eref{momq=1} and are in agreement
   with moments obtained numerically from the exact diagonalization of
   the SYK Hamiltonian.

\begin{figure}[t!]
\centerline{   \includegraphics[width=7cm]{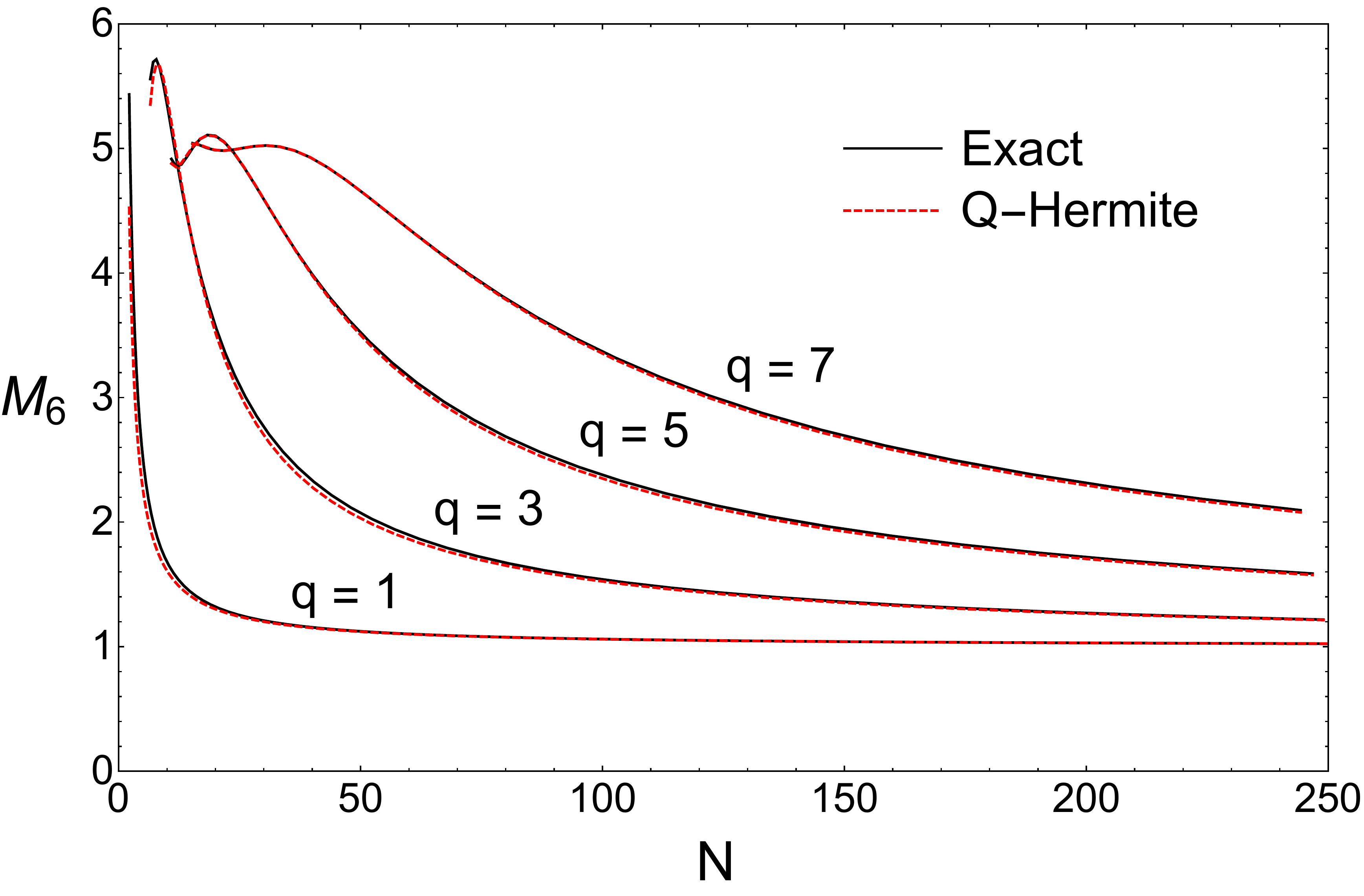}
   \includegraphics[width=7cm]{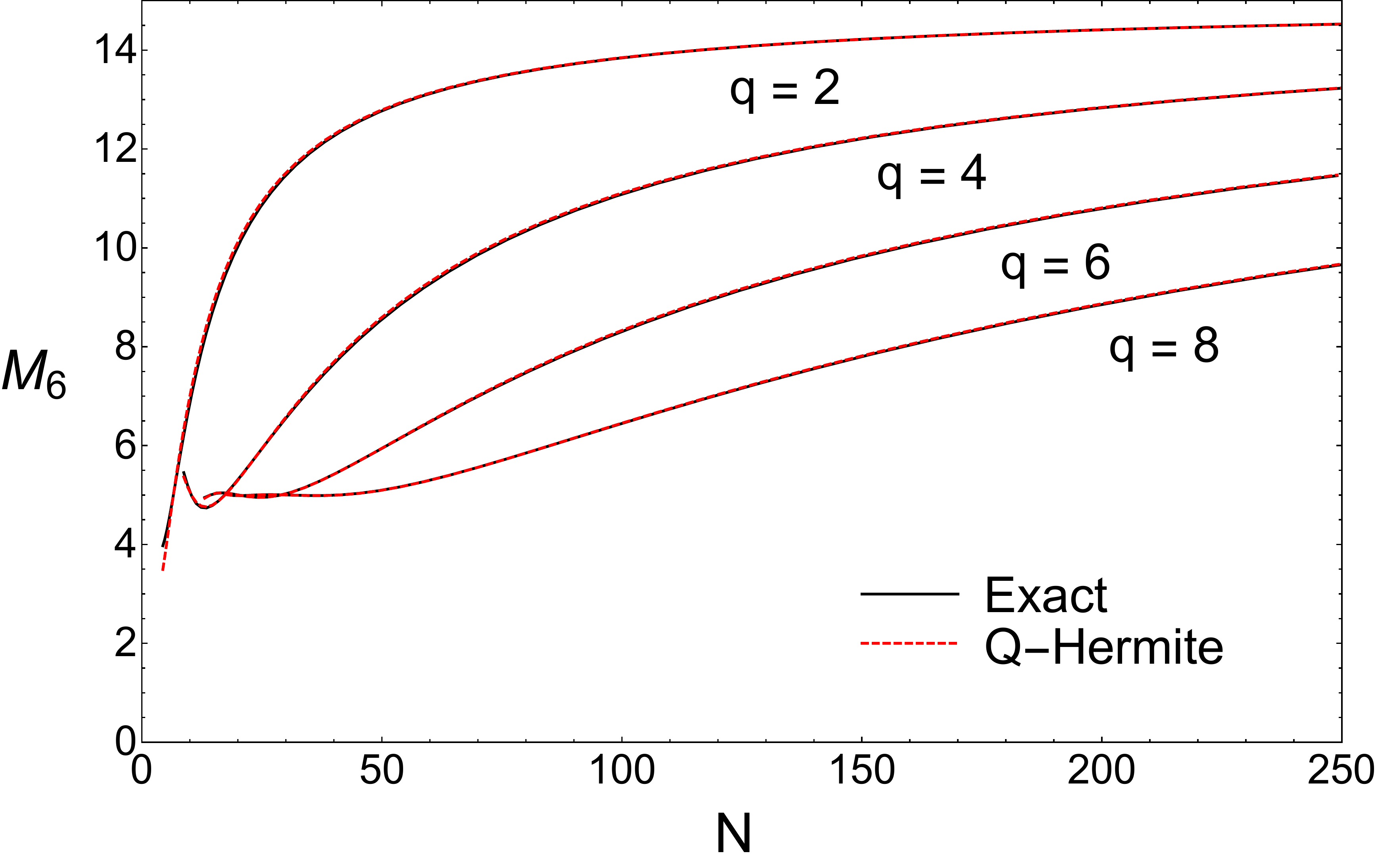}}
   \caption{The $N$-dependence of the sixth moment of the eigenvalue density of the
     SYK model for $q =1,\; 3,\; 5,\; 7$ (left) and  $q =2,\; 4,\; 5,\; 8$ (right). We compare the exact result (solid curve) to the Q-Hermite
     result (dashed).
   }
   \label{fig-m6}
   \end{figure}

\begin{figure}[t!]
\centerline{   \includegraphics[width=7cm]{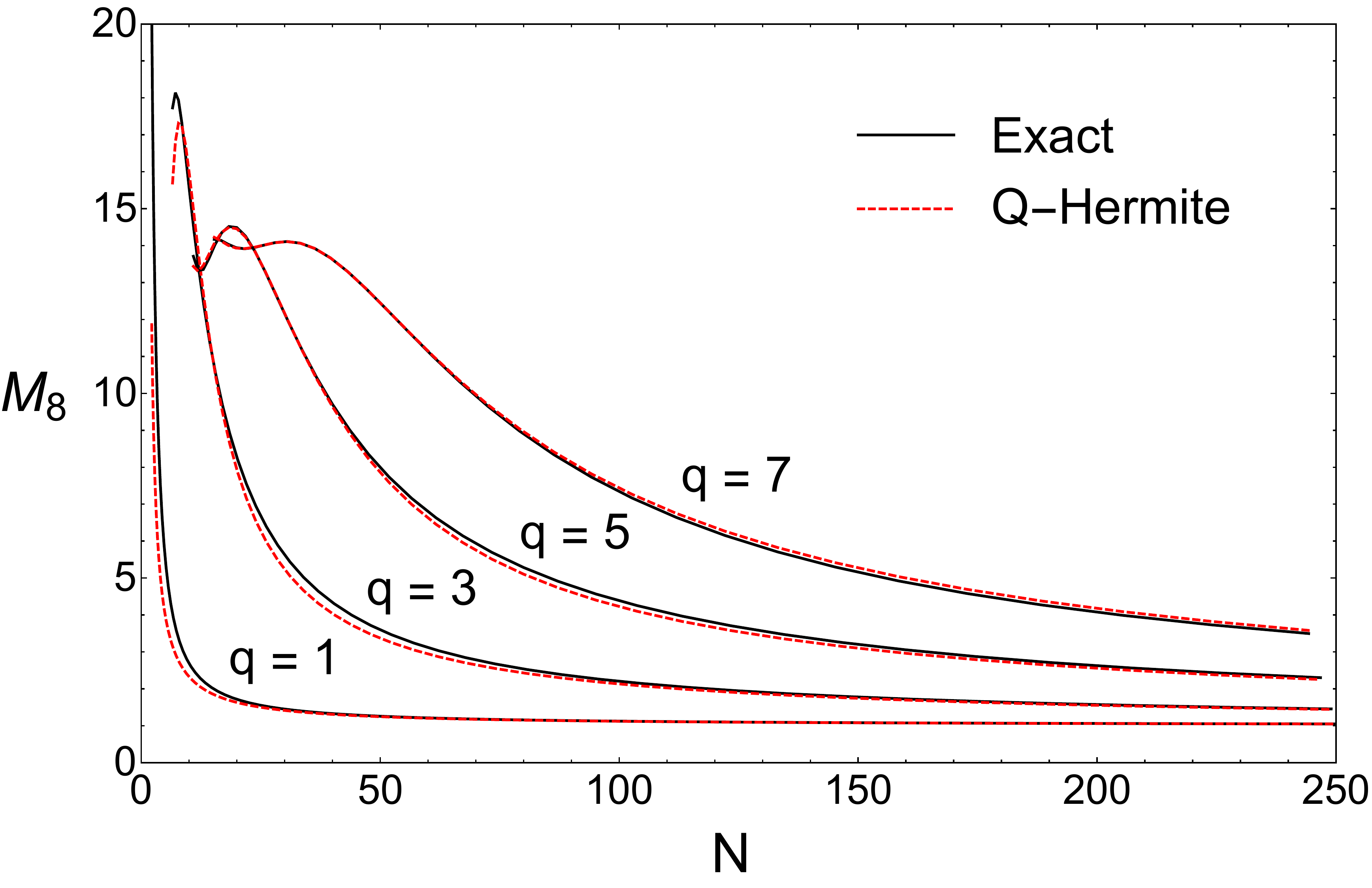}
   \includegraphics[width=7cm]{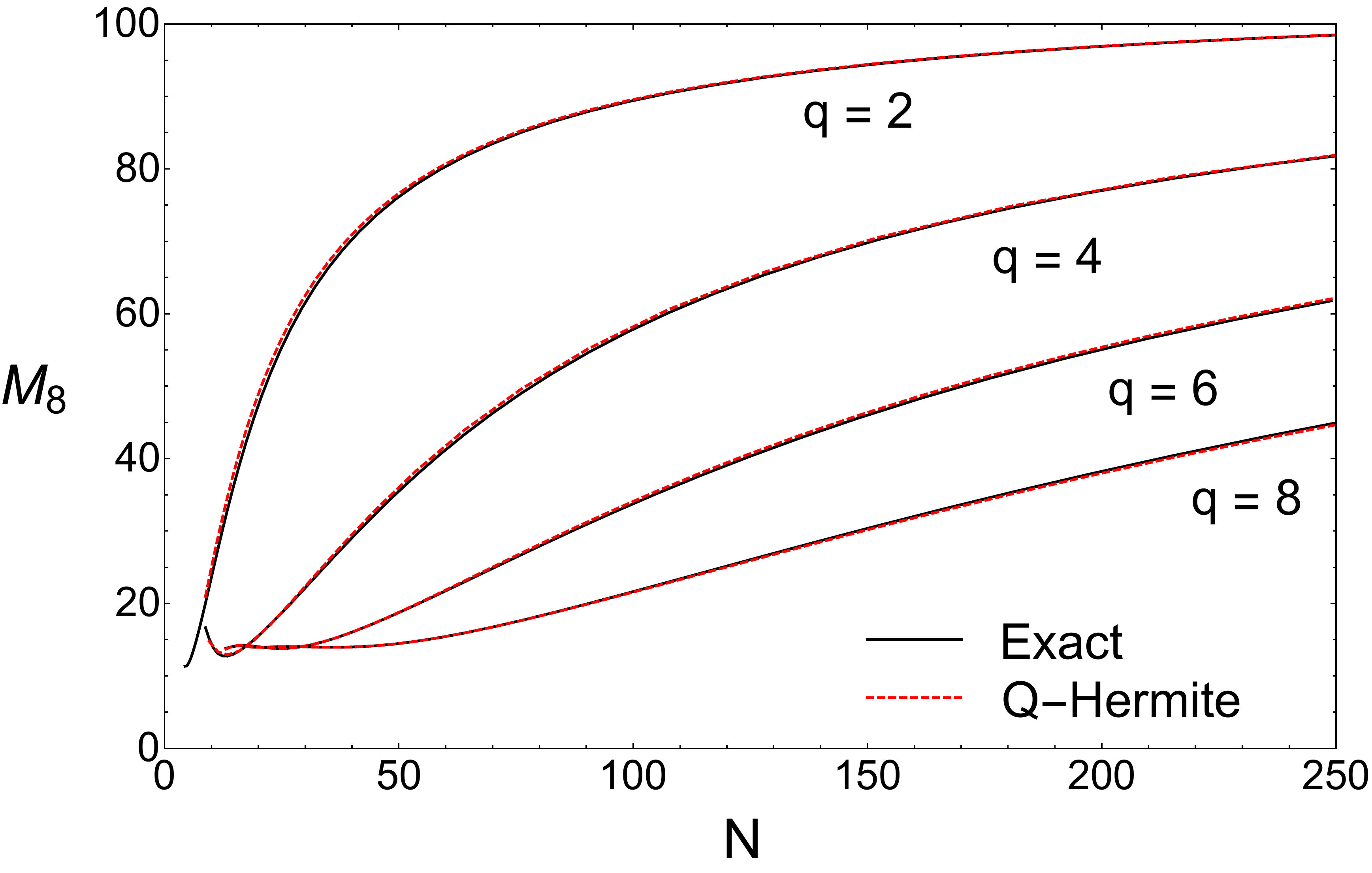}}
   \caption{The $N$-dependence of the eighth moment of the eigenvalue density of the
     SYK model for $q =1,\; 3,\; 5,\; 7$ (left) and  $q =2,\; 4,\; 5,\; 8$ (right).
We compare the exact result (solid curve) to the Q-Hermite
     result (dashed).   }
   \label{fig-m8}
 \end{figure}

   In figure \ref{fig-m6} and figure~\ref{fig-m8} we show the $N$ dependence of
   the sixth and eighth moment, respectively. We compare the exact result
   the Q-Hermite result to $q=1, 2, \cdots, 8$ and find that the two are
   close in particular for even $q$, even for small values of $N$.

\section{The nature of the Q-Hermite approximation and higher order corrections} \label{sec:qHermiteNature}
The results we have obtained are, at least superficially, contradictory. 
The $1/N^2$ correction to the moments of the Q-Hermite approximation, one of the main result of the paper, does not improve the spectral density in a uniform sense.
At the same time the exact computation of low order moments (up to 8th moment),
show that the Q-Hermite approximation gives surprisingly accurate results even for relatively small $N$.
Adding to this point, the $1/N^2$ correction we have obtained in this paper only improves the Q-Hermite for $2p \ll N/q^2$. At small $N$, it actually makes the approximation worse than using Q-Hermite results alone even for relatively small $p$.

All the above suggests that the Q-Hermite approximation should be understood
as a re-summed finite $N$ result, which happens to be $1/N$-exact in the large $N$ expansion, while the extra $1/N^2$ corrections from triangle counting
are strictly asymptotic. However, a resummation that is only
$1/N$ exact does not imply that the Q-Hermite approximation must be so accurate at finite $N$. To achieve further understanding let us expand the Q-Hermite moments in powers of $1/N$,
\begin{align}
\label{eqn:QHtoNsquareEvenq}
  \frac{M_{2p}^{\text{QH, even $q$}}}{M_2^p} = &(2p-1)!!\left\{1-\frac 23 {p\choose 2} \frac {q^2}N
  \right .
  \nn\\
&+  \left .
\left[\frac{1}{45}{p\choose 2}(5p^2-p+12)q^4-\frac{4}{3}{p\choose 2}q^3+
  \frac{2}{3}{p\choose 2}q^2\right]\frac{1}{N^2}
\right\}
+O\left(\frac{1}{N^3}\right),
\end{align}
and
\be\label{eqn:QHtoNsquareOddq}
\frac{M_{2p}^{\text{QH, odd $q$}}}{M_2^p}=1+2{p\choose 2}\frac{q^2}{N} +\left[\frac{p(p-1)^2(p-4)}2q^4+4{p\choose 2}q^3-2{p\choose 2}q^2\right]\frac{1}{N^2}+O\left(\frac{1}{N^3}\right),\hspace*{0.5cm}
\ee
where we have used the expansion of $\eta$ and the identities given in
 appendix \ref{append:edgeCounting} to perform the sums
 $\sum_i (-1)^{qE_i}E_i$ and $\sum_i (-1)^{qE_i}E_i^2$.
Comparing this to the difference 
\be\label{eqn:QHcorrectionNsquareEvenq}
\frac{M_{2p}^{\text{even $q$}}-M_{2p}^{\text{QH, even $q$}}}{M_2^p}= -\frac{8}{15}(2p-1)!!\binom{p}{3}\frac{q^3}{N^2}+O\left(\frac{1}{N^3}\right),
\ee
and
\be\label{eqn:QHcorrectionNsquareOddq}
\frac{M_{2p}^{\text{odd $q$}}-M_{2p}^{\text{QH, odd $q$}}}{M_2^p}=8\binom{p}{3}\frac{q^3}{N^2}+O\left(\frac{1}{N^3}\right),
\ee
we observe that the Q-Hermite approximation is not only exact at order $1/N$,
but also almost exact at order $1/N^2$ in the following sense:
\begin{itemize}
\item The Q-Hermite approximation captures the term $q^4/N^2$, which is the only term
  of order  $1/N^2$ that survives in the scaling limit $ N\to \infty$ with
  $q^2/N=\text{constant}$. We have already seen that this
  property holds to all orders in $q^2/N$.
\item Even for fixed $q$, the Q-Hermite approximation contains the dominant contribution to moments, that is,
  the leading coefficients of $1/N^2$ in (\ref{eqn:QHtoNsquareEvenq})
  and (\ref{eqn:QHtoNsquareOddq}) go as $p^4q^4$,
  while the corrections in (\ref{eqn:QHcorrectionNsquareEvenq})
  and (\ref{eqn:QHcorrectionNsquareOddq}) go as $p^3q^3$, and we have $p^4q^4\gg p^3q^3$ already for relatively small $p$ and $q$. 
\end{itemize}
The second property is also likely to hold to all orders in $1/N$. 

Note that in the truncated form of the Q-Hermite result
(\ref{eqn:QHtoNsquareEvenq}), the corresponding spectral density
at each order is a Gaussian or a  derivative of Gaussian with the
same distribution width, so is the spectral density of the extra
correction (\ref{eqn:QHcorrectionNsquareEvenq}). Therefore the breakdown of
the $1/N$ expansion  for $x=E/E_0 \approx 1$
discussed at the end of section \ref{sec:evenqSpecDen} is really an artefact of
the asymptotic nature of this expansion.

 \section{Conclusions and Outlook} 
 \label{sec:conclusions}
 We have obtained analytically an exact expression for the moments of the $q$-body SYK model. For any $q$, we have computed them explicitly up to $1/N^2$ order.
 One surprising result of the calculation of the $1/N^2$ order is that
 it allows a simple and beautiful geometric interpretation in the form of
 triangular loops of the intersection graphs.
 The $1/N$ corrections which are part of the Q-Hermite approximation are
 also geometric in nature and are
 given by the contribution from the edges.  
 Our results can be generalized to higher orders in $1/N$. Preliminary results for the $1/N^3$ correction to the moments
 indicate
 they are also characterized by the geometry of the intersection graphs in a simple manner.
 From a more mathematical perspective, they generate a remarkable set of
 graph-theoretic identities.
 In particular, the $1/N^3$ computation enumerates a particular linear combination of the last three geometric objects in table \ref{tab:8thMomIntersecGraphs}.
 It is one of the miracles of the SYK model that the $1/N^2$ and even higher order corrections can be calculated analytically. On top of this, given that we have an exact
 expression for all contraction diagrams contributing to the moments,
 this might be an indication that the moment
 problem of the SYK model is completely solvable. In future work
 we hope to elaborate on this question.

 The original motivation to carry out the moments calculation was to obtain a more accurate description of the spectral density and also, closely related, to understand better why the Q-Hermite approximation, which reproduces only the $1/N$ correction to the exact moments of the SYK model, is so close to the numerical SYK spectral density at least for even $q$. Even more remarkable is that the Q-Hermite spectral
 density exactly reproduces the temperature dependence of the free energy of the SYK model to leading order in $1/q^2$ at all temperatures both for even and odd $q$.  For the moment, we have at best partial answers to these questions.
 
 We have found that $1/N^2$ corrections to the moments, though exact, lead to a correction to the Q-Hermite spectral density which is only accurate sufficiently far away from the tail of the spectrum.
 As the spectral edge is approached, it gives unphysical results. This is indeed consistent with the fact that, close to the edge of the spectrum, the density is $1/N$ exact \cite{Stanford:2017thb,Bagrets:2017pwq,Belokurov:2017eit}, so in this spectral region, the $1/N^2$ correction must be an artefact of the
asymptotic expansion in $1/N$.
In order to understand the reason for this unphysical behavior we  first note
that the in large $N$ limit at fixed $q$, $\eta \to (-1)^q$ and the Q-Hermite spectral density, the leading term of the expansion, tends to a Gaussian for even $q$ and to $\delta$-functions for odd $q$. Interestingly,
the $1/N^2$ correction to the spectral density can be expressed in terms of
derivatives of this large $N$ limit of the spectral density
which strongly suggests that, in terms of a nonlinear $\sigma$-model for
the spectral density, the $1/N$ corrections are an expansion about the
trivial saddle point. We expect that the spectral edge of the Q-Hermite
result is given by a non-trivial saddle point of this effective $\sigma$-model.
 Indeed a similar effect is observed in the calculation of $1/N$ corrections to the semi-circle law in random matrix theory \cite{verbaarschot1984}.  
It is not clear to us how to re-sum the asymptotic $1/N$ expansion so that it yields a vanishing $1/N^2$ correction to the density close to the edge of the spectrum.
 In fact, we would need an expansion
 to all orders in $1/N$ to do that.
An issue that further complicates the solution of this problem is the non-commutativity of the large $p$ (order of the moment) and the large $N$ limit. 
 The main contribution for moments of order $p\gg N/q^2$ comes from the edge of the spectrum, while when
 we first take the large $N$ limit the main contribution to the moments
 resides in the bulk of the spectrum.  The $1/N^2$ corrections also share
 this nonuniform large $N$ behavior.

 Regarding the reason behind the unexpected close agreement between the Q-Hermite density and the exact spectral density of the SYK model, 
 we have found that the leading $1/N^2$ contribution in $q$ to the $2p$-th moment, which scales as $p^4q^4/N^2$, is actually
 included in the Q-Hermite result which helps explain why this approximation,
 with a difference from the exact result that is
 also subleading in $p$,
 goes beyond its natural limit of applicability.
 However, this does not help explain why the full exact $1/N^2$ correction to the density gives worse results than the
 Q-Hermite approach such as spurious $1/N^2$ corrections to the density close to the ground state.
  In future work we plan to address some of these problems.

  \acknowledgments{Y.J. thanks Weicheng Ye for a discussion of some of the combinatorics used in section \ref{sec:exact} and Mario Kieburg is thanked for pointing out the analytical
    result for $q=1$.
  Y.J. and J.V. acknowledge partial support from  U.S. DOE Grant  No. DE-FAG-88FR40388.
  Part of this work was completed at  the Kavli Institute of UC Santa Barbara
  and J.V.  thanks them for the hospitality. In particular, Arkady Vainshtein
  and Gerald Dunne are thanked for
  useful discussion there.
}

\newpage
\appendix

\section{Calculation of the $1/N^2$  corrections}
\label{proof:master}

The starting point is the general formula (\ref{eqn:etaGdCombinatorics}) for $\eta_G$, which we replicate here for easier access of reading:

\begin{equation}\label{etaapp}
\begin{split}
(-1)^{qE}\eta_G &= \binom{N}{q}^{-V}
   \sum\limits_{\{d_{\alpha_k\alpha_l}\}}
   \sum\limits_{\{ d_{\alpha_k\alpha_l\alpha_m}\}}
   \cdots\sum\limits_{\{d_{\alpha_1\ldots\alpha_{V}}\}}
   (-1)^{c(G)} \frac{N!}{(N-Vq+d_2+2d_3+3d_4+\cdots)!} \\
   & \times\prod_{k=1}^{V}\frac{1}{(q-d_{\alpha_k*}-d_{\alpha_k**})!}
   \prod_{1\leq i<j\leq V}\frac{1}{d_{\alpha_i\alpha_j}!}
   \prod_{1\leq i<j<k\leq V}\frac{1}{d_{\alpha_i\alpha_j\alpha_k}!}
   \cdots \frac{1}{d_{\alpha_1\alpha_2\cdots\alpha_{V}}!}.
   \end{split}
\end{equation}

We already argued in eq.~\eref{eqn:1/Ncontroll} that the orders in $1/N$ is controlled by the term 
\[
{N\choose q}^{-V}\frac{N!}{(N-Vq+d_2+2d_3+\cdots)!} \sim N^{-d_2-2d_3-\cdots}.
\]
Note that because of cancellations, the sum \eref{etaapp} may be of higher
order in $1/N$, for example if a diagram contains nested contractions.
Hence the following four cases  contribute to the order of $1/N^2$:
\begin{itemize}
\item \begin{center}$d_2=0, \quad d_{k\ge 3}=0$;\end{center}
\item \begin{center}$d_2=1,\quad  d_{k\ge 3}=0$;\end{center}
\item \begin{center}$d_2=2,\quad d_{k\ge 3}=0$;\end{center}
\item \begin{center}$d_3=1,\quad d_{k\neq 3}=0$.\end{center}
\end{itemize}
We will compute each of the cases and sum them up. Note the summation indices in (\ref{eqn:etaGdCombinatorics}) are $d_{\alpha_{i_1}\cdots\alpha_{i_k}}$, which involves all the $k$-vertex structures in an intersection graph $G$. For example, the sum over $d_{\alpha_{i_1}\alpha_{i_2}}$ involves summing over all edges that connect
arbitrary two vertices in $G$, and there are $V(V-1)/2$ such edges. Hence, the edges we need to sum over are more than just the edges of $G$ itself, and to aid the forthcoming computation, we complete the graph $G$ by adding dashed lines
between all vertices that are not connected by a solid line. In graph theory
this is known as the completion of a graph.
In figure~\ref{fig:graphCompletion} we illustrate two examples of such graph completion.
\begin{figure}
\begin{center}
\begin{tikzpicture}
\draw[fill=black] (0,0) circle (2pt);
\draw[fill=black] (2,0) circle (2pt);
\draw[fill=black] (1,1.73) circle (2pt);
\node at (3.5,0.85) {$\longrightarrow$};
\draw[fill=black] (5,0) circle (2pt);
\draw[fill=black] (7,0) circle (2pt);
\draw[fill=black] (6,1.73) circle (2pt);

\draw[] (2,0) -- (1,1.73);
\draw[]  (7,0) -- (6,1.73);
\draw[thick,red,dashed](7,0) -- (5,0) --(6,1.73);

\draw[fill=black] (0,-3) circle (2pt);
\draw[fill=black] (1.8,-3) circle (2pt);
\draw[fill=black] (0,-1.2) circle (2pt);
\draw[fill=black] (1.8,-1.2) circle (2pt);
\node at (3.5,-2.2) {$\longrightarrow$};
\draw[fill=black] (5,-3) circle (2pt);
\draw[fill=black] (6.8,-3) circle (2pt);
\draw[fill=black] (5,-1.2) circle (2pt);
\draw[fill=black] (6.8,-1.2) circle (2pt);

\draw[] (1.8,-3) -- (0,-3) -- (0,-1.2);
\draw[] (6.8,-3) -- (5,-3) -- (5,-1.2);
\draw[thick,red,dashed](5,-1.2) -- (6.8,-1.2) -- (6.8,-3) --(5,-1.2);
\draw[thick,red,dashed](6.8,-1.2) -- (5,-3);
\end{tikzpicture}
\end{center} 
\caption{Graph completion for a 3-vertex graph and 4-vertex graph. The resulting graphs have edges between each pair of vertices. 
This is equivalent to the edge-2 coloring (i.e. that we use two different colors to color a graph) of complete graphs (i.e. graphs where each pair of vertices is connected by an edge).}\label{fig:graphCompletion}
\end{figure}
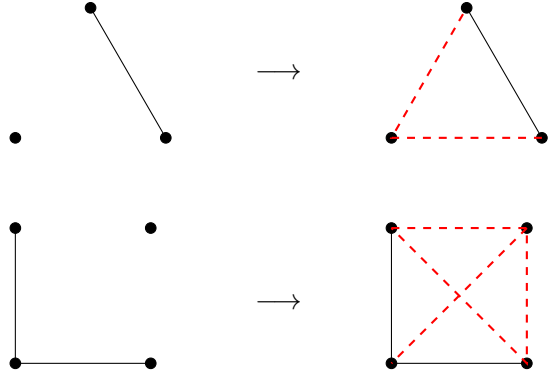 For a completed graph, we denote by
 $w_0, w_1, w_2$ the numbers of wedges with 0, 1 and 2 solid lines, and by $n_0, n_1, n_2, n_3$ the numbers of triangles with 0, 1, 2 and 3 solid lines (see table~\ref{tab:wedgeandTrianNumbers}), which will be useful for the computations to come.
\begin{table}
\begin{center}
\begin{tabular}{|c|c|c|c|c|c|c|c|c|}
\hline \begin{tikzpicture} \node at (0,0) {Structure};\end{tikzpicture}  &\begin{tikzpicture}
\draw[fill=black] (0,0) circle (1pt);\draw[fill=black] (0.8,0) circle (1pt);
\draw[fill=black] (0.4,0.692) circle (1pt);
\node at  (0.4,0.71) {}; \draw[thick,red,dashed](0,0)--(0.4,0.692)--(0.8,0);\end{tikzpicture}& \begin{tikzpicture}
\draw[fill=black] (0,0) circle (1pt);\draw[fill=black] (0.8,0) circle (1pt);
\draw[fill=black] (0.4,0.692) circle (1pt);
\node at  (0.4,0.71) {}; \draw[thick,red,dashed](0,0)--(0.4,0.692);\draw (0.4,0.692)--(0.8,0);\end{tikzpicture} &\begin{tikzpicture}
\draw[fill=black] (0,0) circle (1pt);\draw[fill=black] (0.8,0) circle (1pt);
\draw[fill=black] (0.4,0.692) circle (1pt);
\node at  (0.4,0.71) {}; \draw (0,0)--(0.4,0.692)--(0.8,0);\end{tikzpicture}&\begin{tikzpicture}
\draw[fill=black] (0,0) circle (1pt);\draw[fill=black] (0.8,0) circle (1pt);
\draw[fill=black] (0.4,0.692) circle (1pt);
\draw[thick,red,dashed] (0,0)--(0.4,0.692)--(0.8,0)--(0,0);\end{tikzpicture} & \begin{tikzpicture}
\draw[fill=black] (0,0) circle (1pt);\draw[fill=black] (0.8,0) circle (1pt);
\draw[fill=black] (0.4,0.692) circle (1pt);
\draw[thick,red,dashed] (0,0)--(0.4,0.692)--(0.8,0);\draw (0.8,0)--(0,0);\end{tikzpicture}& \begin{tikzpicture}
\draw[fill=black] (0,0) circle (1pt);\draw[fill=black] (0.8,0) circle (1pt);
\draw[fill=black] (0.4,0.692) circle (1pt);
\draw (0,0)--(0.4,0.692)--(0.8,0);\draw[thick,red,dashed] (0.8,0)--(0,0);\end{tikzpicture} &\begin{tikzpicture}
\draw[fill=black] (0,0) circle (1pt);\draw[fill=black] (0.8,0) circle (1pt);
\draw[fill=black] (0.4,0.692) circle (1pt);
\draw (0,0)--(0.4,0.692)--(0.8,0)--(0,0);\end{tikzpicture}\\ \hline \begin{tikzpicture} \node at (0,0) {number};\node at (0,0.3) {};\end{tikzpicture} & \begin{tikzpicture} \node at (0,0) {$w_0$};\end{tikzpicture} & \begin{tikzpicture} \node at (0,0) {$w_1$};\end{tikzpicture} & \begin{tikzpicture} \node at (0,0) {$w_2$};\end{tikzpicture} & \begin{tikzpicture} \node at (0,0) {$n_0$};\end{tikzpicture} & \begin{tikzpicture} \node at (0,0) {$n_1$};\end{tikzpicture}   & \begin{tikzpicture} \node at (0,0) {$n_2$};\end{tikzpicture}   & \begin{tikzpicture} \node at (0,0) {$n_3$};\end{tikzpicture}  \\ \hline
\end{tabular}
\end{center}
\caption{Definition of $w_0, w_1, w_2$ and $n_0, n_1, n_2, n_3$.}\label{tab:wedgeandTrianNumbers}
\end{table}

\subsection{Cases with $d_{k\ge 3}=0$}

In this case the general expression for contractions reduces to
\be
  \label{d30start}
  (-1)^{qE}\eta_G&=&\binom{N}{q}^{-V}
    \sum\limits_{\{d_{\alpha_k\alpha_l}\}}
  (-1)^{c(G)}
    \frac{N!}{(N-Vq+d_2)!}
    \nn\\ && \times
     \prod_{k=1}^V\frac{1}{(q-d_{\alpha_k*}-d_{\alpha_k**}-\cdots)!}\prod_{1\leq i<j\leq V}\frac{1}{d_{\alpha_i\alpha_j}!},
\ee
     where 
\be
c(G)=\sum_{k=1}^{E} c_{\alpha_{i_k}\alpha_{j_k}}=\sum_{k=1}^{E} d_{\alpha_{i_k}\alpha_{j_k}}.
\label{cg}
\ee
Note the second equality of (\ref{cg}) holds because we are working with the particular case $d_{k\ge 3}=0$. To order $1/N^2$ we have three cases, $d_2=0$, $d_2=1$ and $d_2=2$ which we
will analyze next.

\noindent
{\it Case $d_2=0$.}
The result \eref{d30start} for a contraction diagram simplifies to
\begin{align}\label{eqn:d2=0}
(-1)^{qE}\eta_G^{d_2=0} =&
  \frac{N! (N-q)!^{V}}{(N-Vq)!N!^{V}}\nn \\ =& 1 -\frac{q^2 }N{V \choose 2}
  +\frac 1{2}\frac {q^2}{N^2}{V\choose 2}-\frac 13 \frac{q^3}{N^2}
  {V\choose 2}(V+1) +\frac 12 \frac{q^4}{N^2}{V\choose 2}^2 +O(1/N^3).
\end{align}

\noindent
    {\it Case $d_2=1$.}
  In this case, among the $\binom{V}{2}$ possible $d_{\alpha_k\alpha_l}$ we sum over, 
  $E$ of them occur  in the set of edges of $G$ (solid lines in a completed graph) and they give $c(G) =1$, while ${V \choose 2} - E $ of them are not in this set of edges (dashed lines in the completed graph)
    and they give $c(G) =0$. So this case
contributes to $\eta_G$ by the following expression:

\begin{align}\label{eqn:d2=1}
  (-1)^{Eq}\eta_G^{d_2=1}=&
\frac{N! (N-q)!^{V}}{(N-Vq)!N!^{V}}
\frac {q^2}{N-V q+1} \left( {V\choose 2 } -2 E \right )  \nn\\
=&\left[  \frac {q^2}N + \frac {q^2}{N^2}\left (V q -1 -q^2{V \choose 2}  \right )    \right ]\left( {V\choose 2 } -2 E \right )+O(1/N^3).
\end{align}

\noindent
{\it Case $d_2=2$.}
When one of the $d_{\alpha_k\alpha_l} = 2$ then either $c(G) = 2$ or $c(G)=0$ and in both cases the phase factor is equal to 1.
This results in the contribution
\begin{equation}
  (-1)^{Eq}\eta_G={V \choose 2}
\frac{N! (N-q)!^{V}}{(N-Vq)!N!^{V}}
\frac{q^2(q-1)^2}{2(N-V q+2)(N-V q+1)}
= {V \choose 2}\frac{q^2(q-1)^2}{2N^2}+O(1/N^3),
\label{eqn:d2=2a}
\end{equation}

The case with two of the $d_{\alpha_k\alpha_l}$ equal to 1 is more complicated
because we have to distinguish the case where they share a common index and
the case where they do not. If they do not share a common index
the combinatorial factor apart from the phase factor is
\be
\frac{N! (N-q)!^{V}}{(N-Vq)!N!^{V}}
\frac{q^4}{(N-V q+2)(N-V q+1)}=\frac{q^4}{N^2} +O(1/N^3).
 \ee
 while when they share a common index we obtain
 \be
 \label{eqn:wedgeContribution}
\frac{N! (N-q)!^{V}}{(N-Vq)!N!^{V}}
\frac{q^3(q-1)}{(N-V q+2)(N-V q+1)}=\frac{q^3(q-1)}{N^2} +O(1/N^3).
 \ee
 The coefficient of the $q^4$ term is the same in both cases which
 simplifies the counting. We have ${E \choose 2}$ pairs of indices with $c(G)=2$,
 $E({V \choose 2}-E)$ pairs with $c(G) =1$ and 
${ V(V-1)/2-E \choose 2}$ pairs with $c(G)=0$.
 Summing over the $\{d_{\alpha_k \alpha_l}\}$ also including pairs that share
 a common index, adding the $-q^3/N^2$ contribution from (\ref{eqn:wedgeContribution}) which was not accounted for in the previous counting,
 we obtain to order $1/N^2$
 \be
  (-1)^{Eq}\eta_G^{d_2=2}=& \left [ {E\choose 2}-E \left({V \choose 2}-E \right ) +
       { V(V-1)/2-E \choose 2}  \right ]\frac {q^4}{N^2}
   +   (-1)^{Eq}\eta_G^{w},
\label{eqn:d2=2}
       \ee
       where we have separated the contribution of pairs that share a common
       index,
       \be
       (-1)^{Eq}\eta_G^{w} =  -\frac{q^3}{N^2}(w_0-w_1+w_2),
       \label{etaw}
       \ee
       because it combines naturally with the contributions that
       will be discussed in the next subsection.
The $w_0, w_1$ and $w_2$ are defined in table ~\ref{tab:wedgeandTrianNumbers}, they will give $c(G)=0, 1, 2$, respectively, and hence the signs in (\ref{etaw}).

Summing  all the terms with $ d_{k\ge 3} =0$, i.e. eqs. 
(\ref{eqn:d2=0}), (\ref{eqn:d2=1}) and (\ref{eqn:d2=2}),
the result simplifies to
\begin{equation}\label{eqn:eta2}
  (-1)^{Eq}\eta_G^{d_{k\ge 3}=0} = 1  -\frac{2Eq^2}{N}+\bigg\{ 2Eq^2
  + \Big[ -2EV+2 {V \choose 3} \Big]q^3  +2E^2q^4\bigg\}\frac{1}{N^2} + \frac{q^3}{N^2}(-w_0+w_1-w_2).
\end{equation}

\subsection{Cases with $d_3=1$}

When $d_3 =1$ to order $1/N^2$ only the $d_2 = 0$ terms contribute.
The result for
a contraction diagram is then given by
\be
\label{eqn:d3=1a}
  (-1)^{Eq}\eta_G^{d_3=1}&=&\frac{N! (N-q)!^{V}}{(N-Vq)!N!^{V}}
   \sum\limits_{\{ d_{\alpha_k\alpha_l\alpha_m}\}}
   (-1)^{c(G)}\frac {q^3}{(N-Vq +2)  (N-Vq +1)}
\nn\\
   &=&
   \sum\limits_{\{ d_{\alpha_k\alpha_l\alpha_m}\}}
   (-1)^{c(G)}\frac {q^3}{N^2} +O(1/N^3),\nn\\
\ee
where 
\be\label{eqn:cGFord3=1a}
c(G)=\sum_{k=1}^{E} c_{\alpha_{i_k}\alpha_{j_k}}=\sum_{k=1}^{E} d_{\alpha_{i_k}\alpha_{j_k}*}.
\ee
The second equality \eqref{eqn:cGFord3=1a} is true because we are working in the particular case where $d_{k\ne 3}=0$, and we remind the reader that there is another summation implied by the ``$*$'' in the subscript.   Either 0, 1, 2 or 3 of the edges of $d_{\alpha_k\alpha_l\alpha_m}$ can
   be part of the edges that occur in $c(G)$, and their total
   number are  $n_0$, $n_1$, $n_2$ or $n_3$ respectively, as defined in table \ref{tab:wedgeandTrianNumbers}.
   We thus obtain the contribution
   \be
(-1)^{Eq}\eta_G^{d_3=1}&=&   \frac {q^3}{N^2}(n_0-n_1+n_2-n_3).
\ee
Including the contribution of the wedges in eq. \eref{etaw}, the total $q^3/N^2$ contribution
is given by
\be
(-1)^{Eq}(\eta_G^{d_3=1}+\eta_G^{w})=\frac {q^3}{N^2}(-w_0+w_1-w_2+n_0-n_1+n_2-n_3).
\ee
From the graphical interpretation of these quantities (see Table~\ref{tab:wedgeandTrianNumbers}),  it is clear that
\be
w_0=3n_0+n_1,\qquad w_1=2n_1+2n_2,\qquad w_2=n_2+3n_3.
\ee
The reason is that each wedge is contained in one and only one triangle, and the identities follow by counting in each type of triangle how many wedges of different types occur.
 This
results in the simplification
\be
(-1)^{Eq}(\eta_G^{d_3=1}+\eta_G^{w})=\frac {q^3}{N^2}(-2n_0+2n_2 -4n_3).
\ee
We have the obvious identities \cite{goodman1959}
\be
 n_1 +2n_2 + 3 n_3 &=&(V-2)E,\\
 n_0+n_1+n_2+n_3 &=& {V \choose 3}.
 \ee
 The first identity can be seen as follows. If we have an edge, it can be combined with
 0, 1, or 2 other edges to from a triangle with one solid edge, two solid edges, or three
 solid edges, respectively. In total there are $E(V-2)$ possibilities to combine any
 edge with the remaining vertices, which gives the
 right hand side of this identity.
 The left-hand side counts the same thing triangle by triangle. That is, a triangle with one
 solid edge occurs once,
 a triangle with two solid edges is counted twice because each of the solid edges can be
 taken as the starting point in $E$. For the same reason
a solid triangle is counted three times in $(V-2)E$, and hence the identity.
 By subtracting the two identities we find
 \be
 -n_0 +n_2 + 2n_3 =(V-2)E -{V \choose 3}.
 \ee
The total contribution of the wedge and triangles can thus be written
as
\be
(-1)^{Eq}(\eta_G^{d_3=1}+\eta_G^{w})=\frac {q^3}{N^2} \left (-8n_3 +2(V-2)E - 2{V\choose 3}
\right ).
\ee
Adding the thus contribution to the $d_{k\le2}$ contributions (see eq.
\eref{eqn:eta2}) we obtain
\be
(-1)^{Eq}(\eta_G^{d_{{k\le2}}}+\eta_G^{d_3=1})
    = 1  -\frac{2Eq^2}{N}+ 2E \frac{q^2}{N^2}
    -4E \frac{q^3}{N^2}  +2E^2\frac{q^4}{N^2} -
    8n_3 \frac{q^3}{N^2}.
\ee
Comparing this to the Q-Hermite result, the $1/N^2$ correction simplifies
to
\be
(-1)^{Eq}(\eta_G^{d_{{k\le2}}}+\eta_G^{d_3=1})
     =
(-1)^{E q} \eta^{E} 
 -    8n_3 \frac{q^3}{N^2}+O(1/N^3).
 \ee
This proves (\ref{mom1n2}).

  \section{Scaling limit of $\eta$}\label{append:scalingLimitEta}

  To obtain the large $N$ double scaling limit of $\eta$ at fixed $q^2/N$ we
  express $\eta$ in terms of the hypergeometric function as
  \be
  \eta = {N\choose q}^{-1}{N-q \choose q} {\,}_2F_1(-q,-q,N+1-2q,-1) .
  \ee
  The double scaling limit of the binomial factor is given by
  \be
     {N\choose q}^{-1}{N-q \choose q} \approx \left (1- \frac qN \right) ^q
     \sim e^{-q^2/N}.
     \label{bnsym}
     \ee
     The hypergeometric function   $u:={}_2F_1(-q,-q,N+1-2q,z)$ satisfies the
     differential equation
  \be  
  z(1-z) \frac{d^2u}{dz^2} +(N+1-2q+(2q-1)z) \frac{du}{dz} - q^2u=0.
  \ee
  In the double scaling limit this simplifies to
  \be
  N\frac {du}{dz} -q^2u=0,
  \ee
  which is solved by
  \be
  u=c e^{q^2 z/N}.
  \ee
  The constant is fixed to $c=1$ by the requirement that $\eta = 1$ for $z=1$.
  For $z=-1$ we thus obtain the asymptotic double scaling limit
\be
  _2F_1(-q,-q,N+1-2q,-1)\sim e^{-q^2/N},
\ee
and using eq. \eref{bnsym} this results in the scaling limit
\be
\eta \sim e^{-2q^2/N}.
\ee

\section{Edge counting from the Riordan-Touchard formula}
\label{append:edgeCounting}

To evaluate the $1/N$ and $1/N^2$ contributions to the Q-Hermite moments we
need the sums
\be
\sum\limits_{i=1}^{(2p-1)!!}(-1)^{qE_i} E_i \qquad {\rm and } \qquad \sum\limits_{i=1}^{(2p-1)!!}(-1)^{qE_i}{E_i}^2 .
\ee
They follow from the first and second derivatives of the Riordan-Touchard
formula at $\eta =1$
\begin{equation}
  \phi_p(\eta):=\sum\limits_{i=1}^{(2p-1)!!}\eta^{E_i}
  =\frac{1}{(1-\eta)^p}\sum\limits_{k=-p}^{p}(-1)^k\eta^{k(k-1)/2}\binom{2p}{p+k},
\end{equation}
where the sum is over all contractions for the $2p$-th moment, and $E_i$ is
the number of crossing for the $i$-th chord diagram.
Below we will show that
\begin{align}
\phi'_p(1)&=\sum\limits_{i=1}^{(2p-1)!!}{E_i}
= \frac{1}{3}\binom{p}{2}(2p-1)!!, \label{eqn:totEdge}\\
\phi''_p(1)&=\sum\limits_{i=1}^{(2p-1)!!}{E_i}({E_i}-1)
=\frac{1}{90}\binom{p}{2}(5p^2-p-18)(2p-1)!!,\label{eqn:quadraticEdge}\\
-\phi'_p(-1)&=\sum\limits_{i=1}^{(2p-1)!!}(-1)^{{E_i}}{E_i} = -\binom{p}{2}, \label{eqn:gradedtotEdge}\\
\phi''_p(-1)&=\sum\limits_{i=1}^{(2p-1)!!}(-1)^{{E_i}}{E_i}({E_i}-1)
=6 {p\choose 4}.\label{eqn:gradedQuadraticEdge}
\end{align}
The first two equalities were already shown in \cite{flajolet2000}.
Here, we give the key ingredients of the proof. We start from the following integral representation of $\phi_p$,
\begin{equation}
  \phi_p(e^t)=\frac{1}{\sqrt{2\pi}}\int_{-\infty}^{\infty} e^{-x^2/2}x^{2p}H(x,t)^p dx,
  \label{int-rep}
\end{equation}
where 
\begin{equation}
H(x,t) = \frac{2\sinh^2(x\sqrt{t}/2-t/4)}{x^2\exp(t/2)\sinh(t/2)}.
\end{equation}
This can be used to show that $\phi_p(1) = (2p-1)!!$. It also follows
\be
\phi_p(-1) =\frac{1}{\sqrt{2\pi}}\int_{-\infty}^{\infty} e^{-x^2/2} 
(i\sinh(x\sqrt {\pi i} +1)^p dx = 1.
\label{phipmin1}
\ee
To obtain the derivatives of $\phi_p(\eta)$ we expand
$H(x,t)^p$ about $t=0$, 
\begin{equation}
H(x,t)^p = 1-\frac{p\sqrt{t}}{x}+\frac{p(-3+6p-6x^2+x^4)t}{12x^2}+O(t^{3/2}).
\end{equation}
Then
\begin{equation}
\begin{split}
\lim_{\eta\to 1}\phi'_p(\eta)& =\lim_{t\to 0}e^{-t}\frac{d\phi(e^t)}{dt}\\
&= \frac{1}{\sqrt{2\pi}}\int_{-\infty}^{\infty} e^{-x^2/2}x^{2p}\frac{p(-3+6p-6x^2+x^4)}{12x^2} dx\\\
&=\frac{1}{3}\binom{p}{2}(2p-1)!!.
\end{split}
\end{equation}
This proves the first equality (\ref{eqn:totEdge}). The second equality (\ref{eqn:quadraticEdge}) can be similarly proved.

The proof of the last two equalities (\ref{eqn:gradedtotEdge}) and (\ref{eqn:gradedQuadraticEdge}) requires some
more work. Using the integral representation \eref{int-rep} of $\phi_p(\eta)$
we obtain
\begin{align}
\phi'_p(-1) =&\left[e^{-t}\frac{d\phi_p(e^t)}{dt}\right]_{t=i\pi}\nn\\
= &-\frac{1}{\sqrt{2\pi}}\int_{-\infty}^{\infty}
e^{-x^2/2}\left\{p\left[i\sinh(\sqrt{\pi i }x)+1\right]^{p-1} \left(i\cosh(\sqrt{\pi i }x)\right)\left(\frac{x}{2\sqrt{\pi i}}-\frac{1}{2}\right) \right. \nn\\
&\qquad\qquad\qquad\qquad\quad\left.- \frac{p}{2}\left[i\sinh(\sqrt{\pi i }x)+1\right]^{p}\right\}dx
\nn\\
=& -\frac{p}{\sqrt{\pi}}\int_{-\infty}^{\infty} e^{-y^2}
\left\{\left[i\sinh(\sqrt{2\pi i }y)+1\right]^{p-1}
\left(i\cosh(\sqrt{2\pi i }y)\right)
\left(\frac{y}{\sqrt{2\pi i}}-\frac{1}{2}\right)\right.\nn\\
&\qquad\qquad\qquad\qquad\left.- \frac{1}{2}\left[i\sinh(\sqrt{2\pi i }y)+1\right]^{p}\right\}dy,
\end{align}
where we have substituted $x=\sqrt{2}y$ for the last equality. It is straightforward to show
\begin{align}
&I_{1,p} := \frac{1}{\sqrt{\pi}}\int_{-\infty}^{\infty}dy e^{-y^2}\left[i\sinh(\sqrt{2\pi i }y)+1\right]^{p}= 1,\\
&I_{2,p}:=\frac{1}{\sqrt{\pi}}\int_{-\infty}^{\infty}dy e^{-y^2}\left[i\sinh(\sqrt{2\pi i }y)+1\right]^{p}\left(i\cosh(\sqrt{2\pi i }y)\right) =-1,\\
&I_{3,p}:=\frac{1}{\sqrt{\pi}}\int_{-\infty}^{\infty} dye^{-y^2}y\left[i\sinh(\sqrt{2\pi i }y)+1\right]^{p}\left(i\cosh(\sqrt{2\pi i }y)\right) =-\frac{\sqrt{2\pi i}}{2}p.
\end{align}
Hence,
\begin{equation}
\phi'_p(-1)=-p\left(\frac{1}{\sqrt{2\pi i}}I_{3,p-1}-\frac{1}{2}I_{2,p-1}-\frac{1}{2}I_{1,p}\right)=\binom{p}{2},
\end{equation}
which proves (\ref{eqn:gradedtotEdge}).

To prove the last equality, we start with
\begin{equation}
\begin{split}
\phi_p''(-1)=& \left[e^{-t}\frac{d}{dt}\left(e^{-t}\frac{d \phi_p(e^t)}{dt}\right)\right]_{t=i\pi}
= \phi_p'(-1)+\left[\frac{d^2 \phi_p(e^t)}{dt^2}\right]_{t=i\pi}
=\binom{p}{2}+\left.\frac{d^2 \phi_p(e^t)}{dt^2}\right|_{t=i\pi}.
\end{split}
\end{equation}
The second derivative can be expressed in terms of the integral representation
\eref{int-rep} of $\phi_p(\eta)$ as
\begin{equation}
\begin{split}
\left.\frac{d^2 \phi_p(e^t)}{dt^2}\right|_{t=i\pi}=&\frac{1}{\sqrt{2\pi}}\int_{-\infty}^{\infty}dx\ e^{-x^2/2}x^{2p}\left(\partial^2_t[H(x,t)^p]\right)_{t=i\pi} \\
=&\frac{1}{\sqrt{\pi}}\int_{-\infty}^{\infty} dy\  e^{-y^2}(\sqrt{2}y)^{2p}\left(\partial^2_t[H(\sqrt{2}y,t)^p]\right)_{t=i\pi}.
\end{split}
\end{equation}
After some tedious but straightforward manipulations, we get
\begin{equation}
\begin{split}
\left.\frac{d^2 \phi_p(e^t)}{dt^2}\right|_{t=i\pi}=& \left(\frac{p}{4}-\frac{p^2}{2}\right)I_{1,p-1}+\frac{p^2}{2}I_{2,p-1}- \left(\frac{p}{4}-\frac{p^2}{2}\right)I_{1,p}\\
 &+e^{i 3\pi/4}\left(\frac{p}{\sqrt{2\pi}}-\sqrt{\frac{2}{\pi}}p^2\right)I_{4,p-1}+i\left(\frac{e^{i \pi/4}}{\sqrt{2\pi}}p^2-\frac{e^{i 3\pi/4}}{2\sqrt{2}\pi^{3/2}}p\right)I_{3,p-1}\\
 &+\frac{e^{i 3\pi/4}}{\sqrt{2\pi}}p^2 I_{4,p}+\frac{i}{2\pi}(2p^2-p)I_{5,p-1}-\frac{ip^2}{2\pi}I_{5,p}\\
=& \frac{1}{4} p(p-4) (p-1)^2,
\end{split}
\end{equation}
where we have used two additional integrals to obtain the last line.
\begin{align}
&I_{4,p} := \frac{1}{\sqrt{\pi}}\int_{-\infty}^{\infty}dy e^{-y^2}y\left[i\sinh(\sqrt{2\pi i }y)+1\right]^{p}= -\frac{\sqrt{2\pi i}}{2}p,\\
&I_{5,p}:=\frac{1}{\sqrt{\pi}}\int_{-\infty}^{\infty}dy e^{-y^2}y^2\left[i\sinh(\sqrt{2\pi i }y)+1\right]^{p}= \frac{1}{2}-\binom{p}{2}\pi i.
\end{align}
This proves (\ref{eqn:gradedQuadraticEdge}).

\section{Cut-vertices and factorization}
\label{append:CutVertex}
Since the subscript space of $\binom{N}{q}$ elements is isotropic we can always fix one index and the  result of a  diagram does not depend on this index. So summing over this index gives
a factor ${N \choose q}$.

We define a cut-vertex as a vertex that when it is  cut, the graph becomes disconnected.
A graph without any cut-vertex is called a non-separable or a two-connected graph. 
If we apply the reasoning of the first paragraph to a cut vertex, we immediately arrive at the theorem
\begin{theorem}\label{theo:graphFactorization}
If a graph $G$ contains a cut-vertex, which separates $G$ into subgraphs $G_1$ and $G_2$, then
\[
\eta_G=\eta_{G_1}\eta_{G_2}.
\] 
\end{theorem}

 As an example, the following graphs all contain one cut-vertex (drawn in red):
\begin{center}
\begin{tikzpicture}
\draw[fill=black] (-4.5,0) circle (2pt);
\draw[fill=red] (-3,0) circle (3pt);
\draw[fill=black] (-1.5,0) circle (2pt);

\draw[fill=black] (0,0) circle (2pt);
\draw[fill=red] (2,0) circle (3pt);
\draw[fill=black] (1,1.73) circle (2pt);
\draw[fill=black] (3.5,0) circle (2pt);

\draw[fill=black] (5,0) circle (2pt);
\draw[fill=red] (7,0) circle (3pt);
\draw[fill=black] (6,1.73) circle (2pt);
\draw[fill=black] (7,1.73) circle (2pt);
\draw[fill=black] (8.73,1.73) circle (2pt);
\draw[fill=black] (8.73,0) circle (2pt);

\draw[] (-4.5,0) -- (-3,0) -- (-1.5,0);
\draw[] (2,0) -- (1,1.73) -- (0,0) -- (2,0) -- (3.5,0);
\draw[] (7,0) -- (6,1.73) -- (5,0) -- (7,0) -- (7,1.73) -- (8.73,1.73) -- (8.73,0) -- (7,0) ;
\end{tikzpicture}
\end{center}
Theorem \ref{theo:graphFactorization}, for example implies,
\begin{center}
\begin{tikzpicture}
\draw[fill=black] (0,0) circle (2pt);
\draw[fill=red] (2,0) circle (3pt);
\draw[fill=black] (1,1.73) circle (2pt);
\draw[fill=black] (3.5,0) circle (2pt);
\draw[fill=black] (5,0) circle (2pt);
\draw[fill=red] (7,0) circle (3pt);
\draw[fill=black] (6,1.73) circle (2pt);
\draw[fill=red] (8.25,0.85) circle (3pt);
\draw[fill=black] (9.75,0.85) circle (2pt);

\node at (2,-0.4) {$\eta_{G}$};
\node at (6,-0.4) {$\eta_{G_1}$};
\node at (9.3,-0.3) {$\eta_{G_2}(=\eta)$};

\node at (4.25,0.85) {$=$};
\node at (7.5,0.85) {$\times$};
\draw[] (2,0) -- (1,1.73) -- (0,0) -- (2,0) -- (3.5,0);
\draw[] (7,0) -- (6,1.73) -- (5,0) -- (7,0);
\draw[] (8.25,0.85) -- (9.75,0.85);
\end{tikzpicture}
\end{center}

\section{Moments for $q=2$}\label{append:q=2Moments}
In this appendix we  calculate the moments $M_{2p}/M_2^p$ to order $1/N^2$ for
the $q=2$ SYK model.

Since we are interested in large $N$ asymptotics, it is immaterial whether $N$ is even or odd, and for technical simplicity we choose $N$ to be even. 
The Hamiltonian for $q=2$ model is given by
\be H=i\sum_{i<j}J_{ij}\gamma_i\gamma_j.\ee
This can be rewritten as \cite{cotler2016}
\be H=\sum_{k=1}^{N/2}x_k(2c_k^{\dagger}c_k-1),\ee
where $x_k$ are the positive eigenvalues of the antisymmetric matrices $J_{ij}$, and $c_k, c_k^\dagger$ are the annihilation and creation operators for Dirac fermions. We have 
\be\langle\Tr(H^{2p})\rangle= \left\langle\sum_{ \{s_k=\pm 1\}}\left(\sum_{k=1}^{N/2}s_kx_k\right)^{2p}\right\rangle,\ee
where
\be
\sum_{ \{s_k=\pm 1\}}= \sum_{s_1=\pm 1} \sum_{s_2=\pm 1}\cdots  \sum_{s_{N/2}=\pm 1}.
\ee
Thus we can compute the ensemble average
by averaging over
the joint probability distribution of anti-symmetric Hermitian random matrices
\cite{mehta2004,gross2017},
\be P(x_1,\ldots,x_{N/2}) \prod_{l=1}^{N/2}dx_l= c e^{-\sum_k x_k^2}\prod_{i<j}(x_i^2-x_j^2)^2\prod_{l=1}^{N/2}dx_l. \ee
Here $c$ is a normalization constant, which we have chosen such
that  $P(x_1,\ldots,x_{N/2})$ is normalized to unity.
Note that the joint probability distribution has the parity symmetry $x_k \to -x_k$ for each individual variable, so we can we can restrict ourselves to the
configuration with only positive $s_k$ by compensating with an overall factor
$2^{N/2}$. This results in
\be
\langle\Tr(H^{2p})\rangle= 2^{N/2}\left\langle\left(\sum_{k=1}^{N/2}x_k\right)^{2p}\right\rangle.
\ee
In particular 
\be
\langle\Tr(H^{2})\rangle= 2^{N/2}\left\langle\left(\sum_{k=1}^{N/2}x_k\right)^{2}\right\rangle=2^{N/2} \frac{N}{2} \langle x_1^2\rangle ,
\ee
where we have used $\langle x_k\rangle =0$ and $\langle x_1^2\rangle =\langle x_2^2\rangle=\ldots=\langle x_{N/2}^2\rangle $ because of the parity symmetry
and permutation symmetry of  $P(x_1,\ldots,x_{N/2})$. Hence,
\be
\frac{M_{2p}}{M_2^p}=\left.\left\langle\left(\sum_{k=1}^{N/2}x_k\right)^{2p}\right\rangle \middle/ \left(\frac{N}{2} \langle x_1^2\rangle\right)^p \right. .
\ee
To isolate the leading orders in $1/N$ we now need to analyze the terms in 
\be
\label{eqn:q2MomPartition}
\left\langle\left(\sum_{k=1}^{N/2}x_k\right)^{2p}\right\rangle= \sum_{m_1+\cdots +m_{N/2}=p}\frac{(2p)!}{(2m_1)!(2m_2)!\cdots(2m_{N/2})!}\left\langle x_1^{2m_1}x_2^{2m_2}\cdots x_{N/2}^{2m_{N/2}}\right\rangle
\ee
that are leading orders in $N$. Again, because of the permutation symmetry
of $P(x_1,\ldots,x_{N/2})$, the expectation values on the right-hand side of equation (\ref{eqn:q2MomPartition}) only depend on the partition of $p$ into
$\{m_1,m_2,\cdots, m_{N/2}\}$.
Therefore, given a partition with $k$ nonzero $m_i$'s, $\{m_{i_1},m_{i_2},\cdots, m_{i_k}\}$ ($i_1<i_2<\cdots<i_k$ by convention), all the expectation values of the form 
\be
\left\langle x_{j_1}^{2m_{i_1}}x_{j_2}^{2m_{i_2}}\cdots x_{j_k}^{2m_{i_k}}\right\rangle
\ee
have the same value and this results in a multiplicity factor $\binom{N/2}{k}$ from choosing $k$-element subsets of $\{x_1,\ldots,x_{N/2}\}$.  It is not too hard to convince oneself that 
\be
\left.\left\langle x_{j_1}^{2m_{i_1}}x_{j_2}^{2m_{i_2}}\cdots x_{j_k}^{2m_{i_k}}\right\rangle \middle/ \langle x_1^2\rangle ^p \right. \sim O(1),
\ee
so we can isolate the leading terms in $1/N$ from the
binomial factors $\binom{N/2}{k}$, and there can be more multiplicity factors from permuting the $x_j$'s in the same partition, but that does not bring
any factors of $N$. Now it becomes obvious that the leading terms which are relevant to $1/N^2$ accuracy are associated with the largest multiplicity factors $\binom{N/2}{p}$, $\binom{N/2}{p-1}$ and $\binom{N/2}{p-2}$. From equation (\ref{eqn:q2MomPartition}), these are the terms
\be 
M_{2p} &=& \binom{N/2}{p} \frac{(2p)!}{2^p}\left\langle x_1^{2}x_2^{2}\cdots x_{p}^{2}\right\rangle  +\binom{N/2}{p-1}\binom{p-1}{1} \frac{(2p)!}{2^{p-2}4!}\left\langle x_1^{4}x_2^{2}\cdots x_{p-1}^{2}\right\rangle \\
&& +\binom{N/2}{p-2}\left[ \binom{p-2}{2}\frac{(2p)!}{2^{p-4}4!4!}\left\langle x_1^{4}x_2^{4}x_3^2\cdots x_{p-2}^{2}\right\rangle+\binom{p-2}{1}\frac{(2p)!}{2^{p-3}6!}\left\langle x_1^{6}x_2^{2}\cdots x_{p-2}^{2}\right\rangle
  \right],\nn
\ee
where the factors $\binom{p-1}{1}$,$\binom{p-2}{2}$ and $\binom{p-2}{1}$ come from permutations within a partition.
The rescaled moments are given by
\be
\begin{split}
\frac{M_{2p}}{M_2^p}=&\left(\frac{N}{2}\right)^{-p}\binom{N/2}{p} \frac{(2p)!}{2^p}\frac{W_1}{W_0^p}  +\left(\frac{N}{2}\right)^{-p}\binom{N/2}{p-1}\binom{p-1}{1} \frac{(2p)!}{2^{p-2}4!}\frac{W_2}{W_0^p} \\
&+\left(\frac{N}{2}\right)^{-p}\binom{N/2}{p-2}\left[ \binom{p-2}{2}\frac{(2p)!}{2^{p-4}4!4!}\frac{W_3}{W_0^p}+\binom{p-2}{1}\frac{(2p)!}{2^{p-3}6!}\frac{W_4}{W_0^p}
\right].
\end{split}
\ee
where the combinations $W_k$ are defined by
\be\label{eqn:selberg}
\begin{split}
&W_0 := \langle x_1^2\rangle, \\
&W_1 := \left\langle x_1^{2}x_2^{2}\cdots x_{p}^{2}\right\rangle, \\
&W_2 := \left\langle x_1^{4}x_2^{2}\cdots x_{p-1}^{2}\right\rangle, \\
&W_3 :=\left\langle x_1^{4}x_2^{4}x_3^2\cdots x_{p-2}^{2}\right\rangle, \\
&W_4 :=\left\langle x_1^{6}x_2^{2}\cdots x_{p-2}^{2}\right\rangle. \\
\end{split}
\ee
Before evaluating these averages we can already expand the prefactors to
order $1/N^2$,
\be
\label{eqn:q2MomentIntermediate}
 \begin{split}
\frac{M_{2p}}{M_2^p}=&(2p-1)!!\left\{\left[1-2\binom{p}{2}\frac{1}{N}+(3p-1)\binom{p}{3}\frac{1}{N^2}\right]\frac{W_1}{W_0^p}  \right.\\ 
&\left.+\frac{1}{3}\binom{p}{2}\left[\frac{2}{N}-4\binom{p-1}{2}\frac{1}{N^2}\right]\frac{W_2}{W_0^p} 
 +\frac{4}{N^2}\left[ \frac{1}{3}\binom{p}{4}\frac{W_3}{W_0^p}+\frac{1}{15}\binom{p}{3}\frac{W_4}{W_0^p}
\right]\right\}+O\left(1/N^3\right).
\end{split}
\ee
The averages in (\ref{eqn:selberg}) are Selberg-type integrals and they can be evaluated by using the recursive relations developed in
\cite{verbaarschot:1994gr}. After some algebra we obtain
\be
\begin{split}
&W_0 = \frac{N-1}{2}, \\
&W_1 = \prod_{k=0}^{p-1}\left(\frac{N}{2}-p+k+\frac{1}{2}\right), \\
&W_2 =\left(N-p+\frac{3}{2}\right)\prod_{k=0}^{p-2}\left(\frac{N}{2}-p+k+\frac{3}{2}\right), \\
&W_3 =\left(N-p+\frac{3}{2}\right)\left(N-p+\frac{5}{2}\right)\prod_{k=0}^{p-3}\left(\frac{N}{2}-p+k+\frac{5}{2}\right), \\
&W_4 =\left(N+\frac{1}{2}\right)\left(N-p+\frac{5}{2}\right)\prod_{k=0}^{p-3}\left(\frac{N}{2}-p+k+\frac{5}{2}\right)+\left(\frac{N}{2}-p+2\right)\prod_{k=0}^{p-2}\left(\frac{N}{2}-p+k+\frac{3}{2}\right). \\
\end{split}
\ee
Hence, to the relevant orders, we have 
\be
\begin{split}
&\frac{W_1}{W_0^p} = 1-2\binom{p}{2}\frac{1}{N}+\frac{1}{3}\binom{p}{2}(3p^2-7p-4)\frac{1}{N^2}+O\left(\frac{1}{N^3}\right), \\
&\frac{W_2}{W_0^p}  =2-(2p^2-4p-1)\frac{1}{N}+O\left(\frac{1}{N^2}\right), \\
&\frac{W_3}{W_0^p}  =4+O\left(\frac{1}{N}\right), \\
&\frac{W_4}{W_0^p} =5+O\left(\frac{1}{N}\right). \\
\end{split}
\ee
Note that the each leading term is just the number of nested contractions when re-expressing
the $W_k$ in products of $\Tr J^{2k} $ to leading order in $1/N$.
Substituting the above results into (\ref{eqn:q2MomentIntermediate}), we finally arrive at
\be
\frac{M_{2p}}{M_2^p}=(2p-1)!!\left[1-\frac{8}{3}\binom{p}{2}\frac{1}{N}+\frac{8}{9}\binom{p}{2}(2p^2-2p-1)\frac{1}{N^2}\right]+O\left(\frac{1}{N^3}\right).
\ee

 \section{Calculation of the eighth moment}\label{append:gemnom}

 The sixth moment was already discussed in \cite{garcia2016} and in this appendix
 we only quote the final result. For the eighth moment we work out all
 contributions explicitly. Although originally the combinatorics for the
 eighth moment were obtained by inspection, in several cases we also show
 how they can be obtained from the general formula \eref{eqn:etaGdCombinatorics}.

 According to the Riordan-Touchard formula we have
 \be
\sum_{i=1}^{(2p-1)!!} \eta^{E_i} =
 \frac 1{(1-\eta)^p} \sum_{k=-p}^p (-1)^k \eta^{k(k-1)/2}
       {2p \choose p+k}.
       \ee
       For $p=3$ the right hand side is given by
       \be
\sum_{i=1}^{15} \eta^{E_i} = 5+ 6\eta +3\eta^2 +\eta^3.
\ee
Only the $\eta^3$ term deviates from the exact result which is
given by
\be
\frac{M_6}{M_2^3} = 5+ 6\eta +3\eta^2 +T_6,
\ee
where
\be
T_{6}&=&          
{N\choose q}^{-2}
  \sum_{q_1=0}^q \sum_{l=0}^{q}\sum_{m=0}^{l}(-1)^{q-q_1-m}
      {N-2q+q_1\choose q-l} {2q-2q_1 \choose m}{N-q \choose q-q_1}
      {q\choose q_1}{q_1\choose l-m}\nn\\
&=&          
{N\choose q}^{-2}
  \sum_{q_1=0}^q \sum_{m=0}^{q}(-1)^{q-q_1-m}
      {N-2q+2q_1\choose q-m} {2q-2q_1 \choose m}{N-q \choose q-q_1}
      {q\choose q_1}
\label{T6}
      \ee
      and
        \be
\eta =  {N \choose q}^{-1} \sum_{r=0}^q  (-1)^{q-r} {q \choose r} {N-q \choose q-r}.
 \label{comb}
 \ee

 \begin{figure}[t!]
\begin{center}
   \includegraphics[width = 14cm]{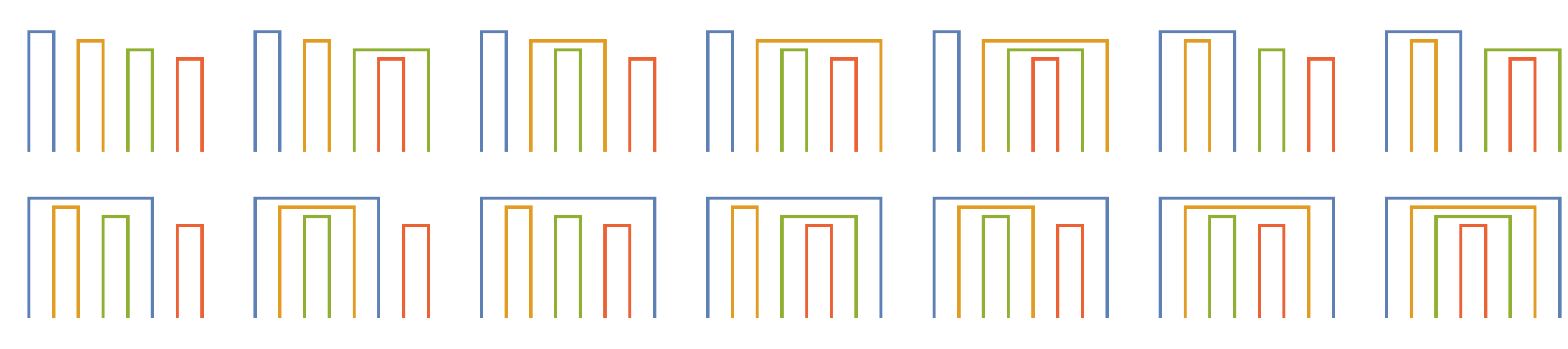}
   \caption{The nested diagrams contributing to the 8th moment.}\label{diak0}
   \end{center}
 \end{figure}

\begin{figure}[t!]
   \begin{center}
   \includegraphics[width = 14cm]{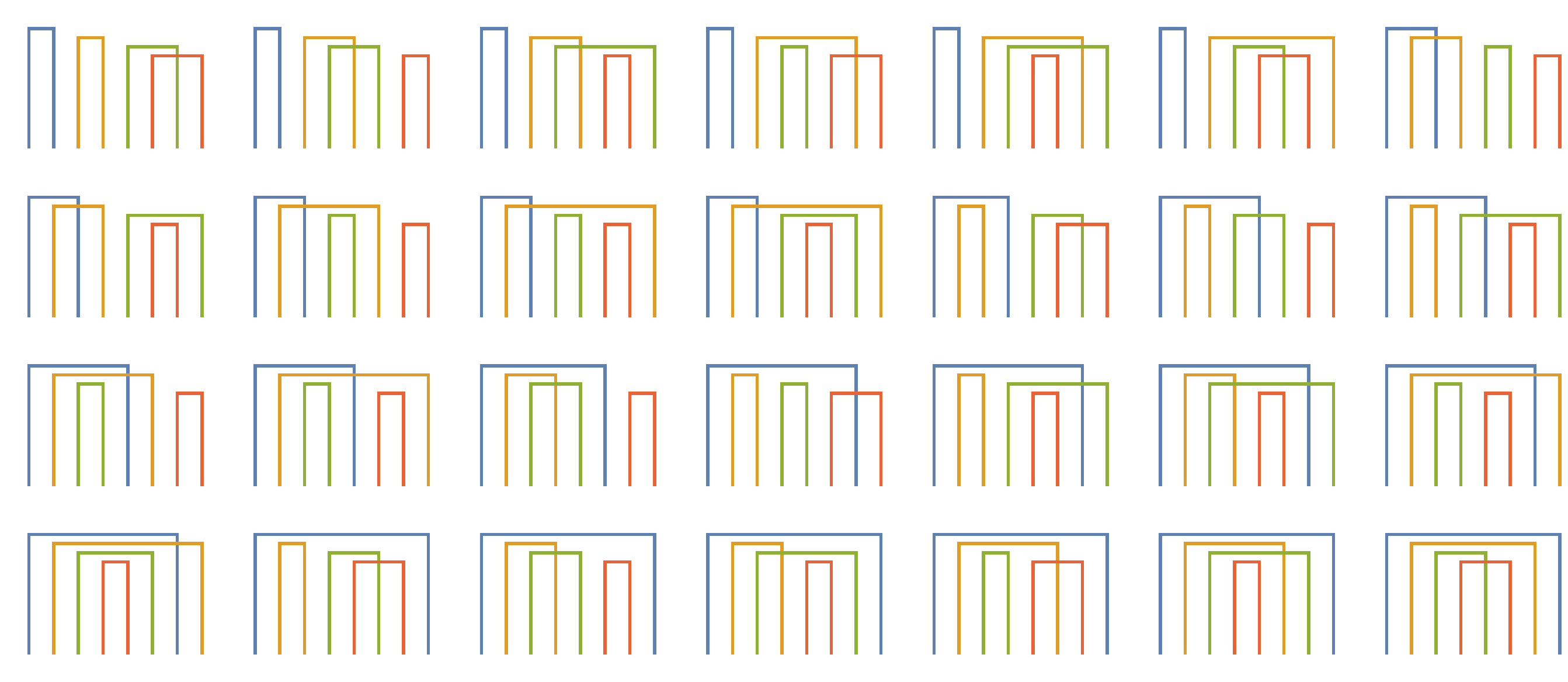}
   \caption{The diagrams with one intersection contributing to the eighth moment.}
   \label{diak1}
   \end{center}
   \end{figure}

\begin{figure}
   \begin{center}
   \includegraphics[width = 14cm]{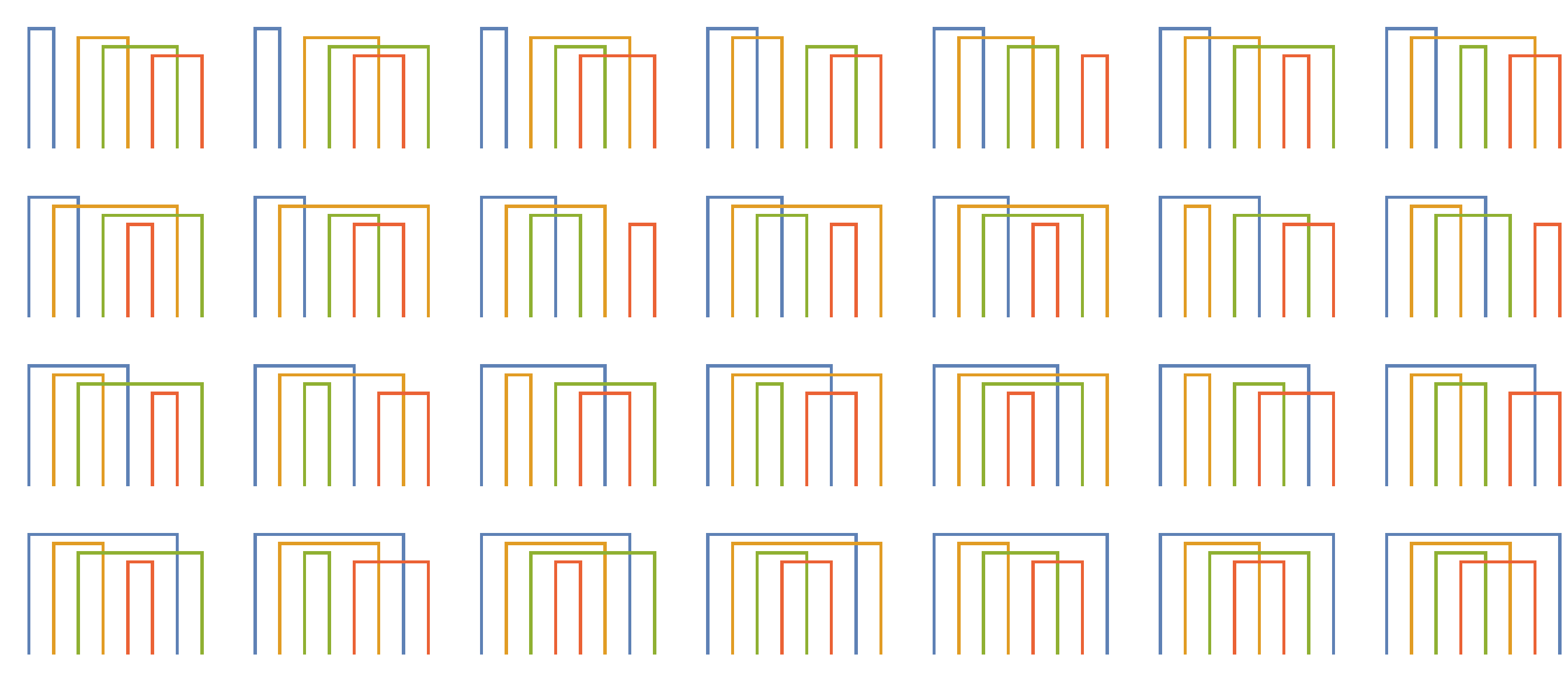}
   \caption{The diagrams with two intersections contributing to the eighth moment.}
   \label{diak2}
   \end{center}
   \end{figure}

 For $p=4$ the Riordan-Touchard formula yields
       \be
       14 + 28 \eta + 28 \eta^2 + 20 \eta^3 +10 \eta^4 + 4\eta^5 +\eta^6 ,
       \ee
   where    the coefficient of $\eta^{E}$ gives the number of diagrams with $E$
       crossings.
       For one and two intersections the crossings can be commuted independently
 resulting in
       \be       
       \frac{M_{8}^{E=0}}{M_2^4} &=& 14, \nn\\
       \frac{M_{8}^{E=1}}{M_2^4} &=& 28 \eta, \label{m8qh}\\
       \frac{M_{8}^{E=2}}{M_2^4} &=& 28 \eta^2,\nn 
       \ee
       The diagrams corresponding to the contributions \eref{m8qh} are shown in
       figures \ref{diak0} to \ref{diak2}.

\begin{figure}
\begin{center}
  \includegraphics[width = 14cm]{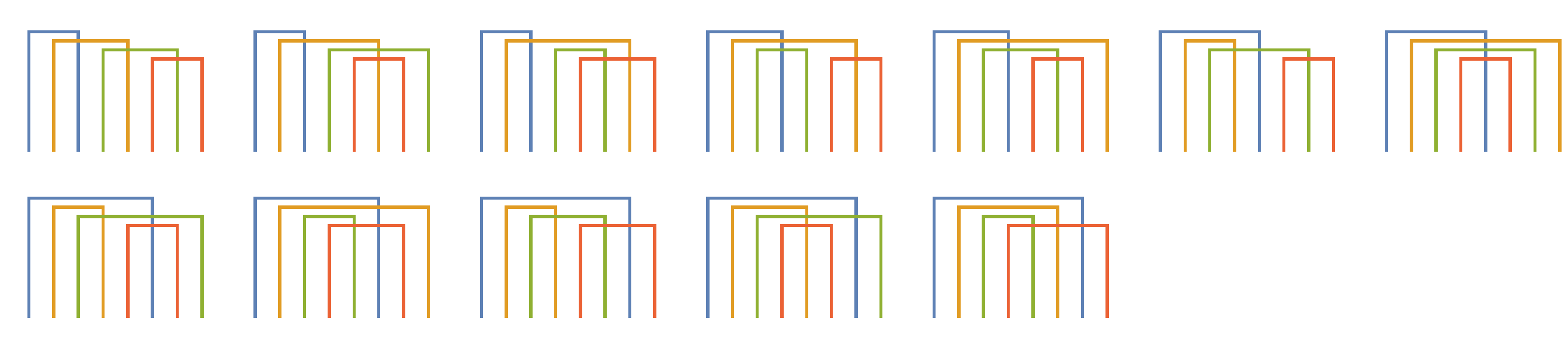}
   \caption{Diagrams with three intersections contributing to the eighth moment  as $\eta^3$.}
   \label{diak3a}
   \end{center}
 \end{figure}

\begin{figure}
  \begin{center}
   \includegraphics[width = 14cm]{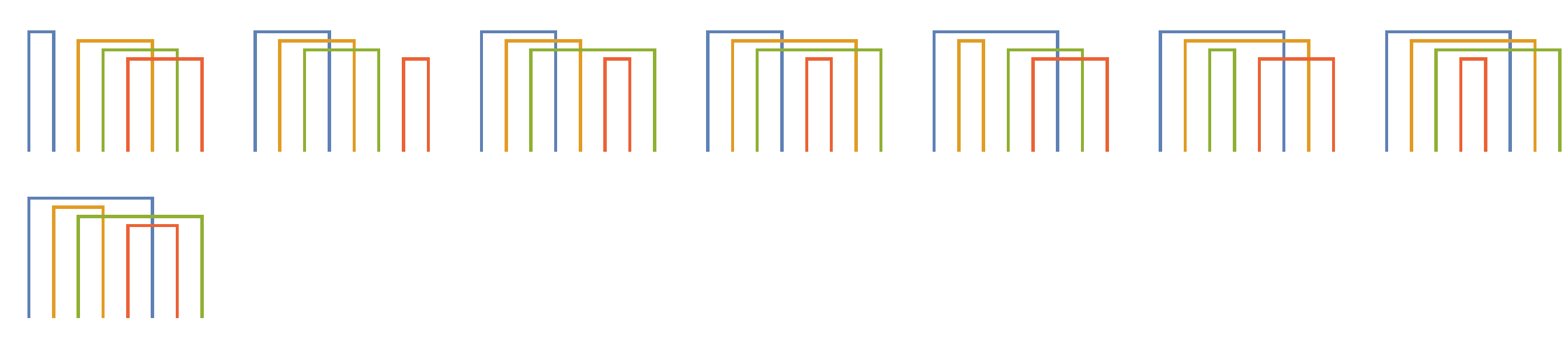}
   \caption{Diagrams with three intersections contributing to the eighth moment  as $T_6$.}
   \label{diak3b}
   \end{center}
 \end{figure}

For $E=3$ we have two different contributions. The first
class of twelve diagrams is shown in figure \ref{diak3a}, where we can move the $\Gamma_\alpha$ to
three consecutive pairs with equal indices by three independent pair exchanges. The results
of each of these diagrams is given by $\eta^3$. The second class of eight diagrams
with three intersections contains a structure we first saw in the calculation
of the sixth moment (see figure \ref{diak3b}). It is given by $T_6$ \cite{garcia2016}, see eq. \eref{T6} and
we thus obtain
\be
       \frac{M_{8}^{E=3}}{M_2^4} &=& 12\eta ^3 + 8 T_6 .
\ee


\begin{figure}[t!]
  \begin{center}
   \includegraphics[width = 4cm]{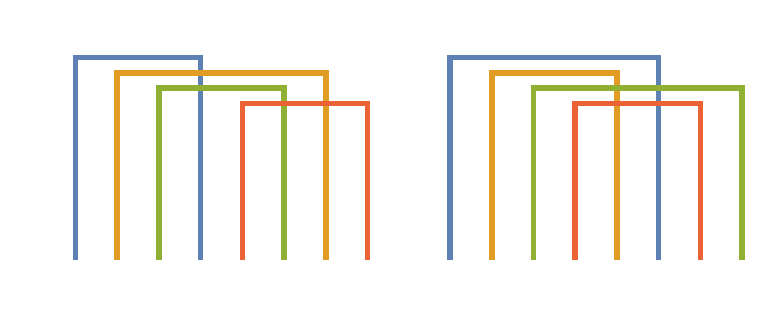}
   \caption{Diagrams with four intersections contributing to the eighth moment  as $T_{44}$.}
   \label{diak4a}
\end{center}
   
 \end{figure}
 
\begin{figure}[t!]
  \begin{center}
   \includegraphics[width = 14cm]{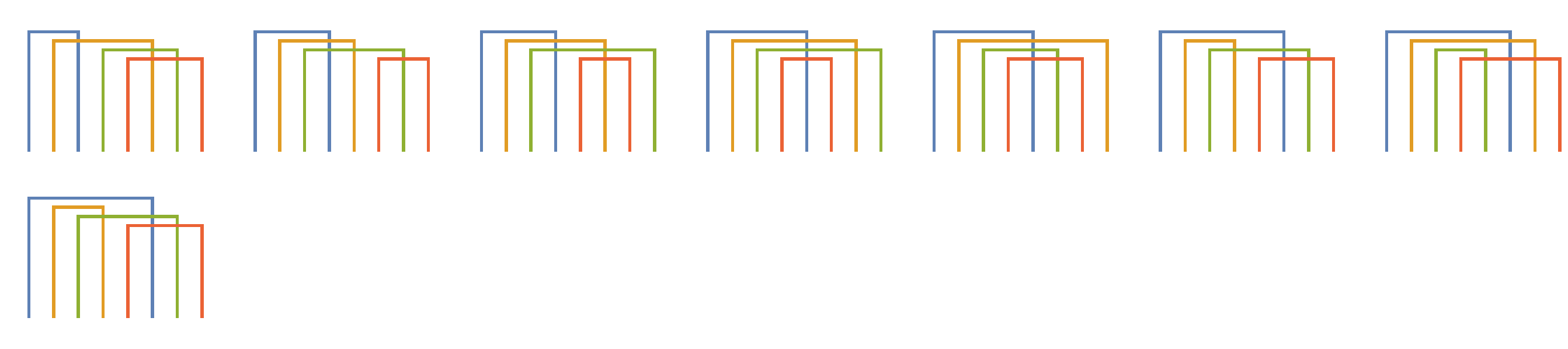}
   \caption{Diagrams with four intersections contributing to the eighth moment as $\eta T_6$.}
   \label{diak4b}
\end{center}
   \end{figure}
\begin{figure}[t!]
  \begin{center}
   \includegraphics[width = 8cm]{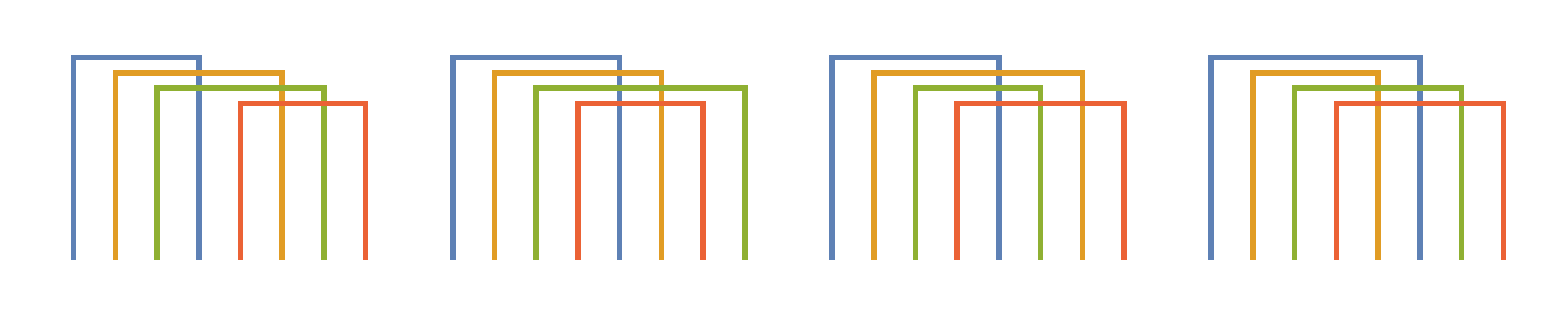}
   \caption{The diagrams with five intersections contributing to the eighth moment
    as $T_{66}$ which is a new type of contribution.}
   \label{diak5}
   \end{center}
   \end{figure}
\begin{figure}[t!]
  \begin{center}
   \includegraphics[width = 2cm]{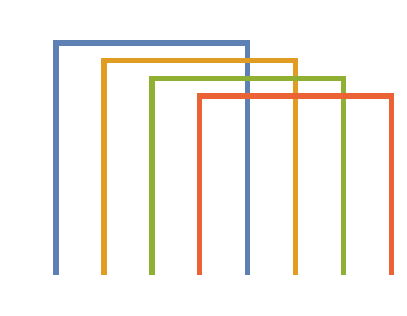}
   \caption{The only diagram with six intersections contributing to the eighth moment. This diagram has not been encountered before and the result will be denoted
     by $T_8$.}
 \label{diak6}
   \end{center}
 \end{figure}
 For $E = 4$ also two different classes of diagrams contribution to the eighth moment. The two diagrams of the first class are shown in figure \ref{diak4a}.
 The result depends on how many gamma matrices the second and third contraction
 have in common. It is given by
 \be
T_{44}&=&
 {N \choose  q}^{-3}\sum_{q_1=0}^q \sum_{l_1=0}^q \sum_{m_1=0}^{l_1} \sum_{l_2=0}^q \sum_{m_2=0}^{l_1} (-1)^{m_1+m_2}
    {N-q \choose q-q_1}{q \choose q_1}
    { N-2q+q_1 \choose q-l_1}
\nn \\ &&\times
    { 2q-2q_1 \choose m_1} {q_1 \choose l_1-m_1}
    { N-2q+q_1 \choose q-l_2}{q_1\choose l_2-m_2}{2q-2q_1  \choose m_2}.
\nn\\ &=&
 {N \choose  q}^{-3}\sum_{q_1=0}^q \sum_{m_1=0}^{q}  \sum_{m_2=0}^{q} (-1)^{m_1+m_2}
    {N-q \choose q-q_1}{q \choose q_1}
    { N-2q+2q_1 \choose q-m_1}
\nn \\ &&\times
    { 2q-2q_1 \choose m_1} 
    { N-2q+2q_1 \choose q-m_2}{2q-2q_1  \choose m_2}.
\ee
  The second class of eight diagrams
 has one intersecting contraction which can be removed by a pair exchange, and
 three other contractions which have three intersections and contribute
 as $T_6$ (see figure \ref{diak4b}). The total contribution of diagrams with four intersections is
 thus given by
 \be
       \frac{M_{8}^{E=4}}{M_2^4} &=& 2 T_{44} + 8 \eta T_6. 
 \ee
There are four diagrams with 5 intersections which are shown in figure \ref{diak5}. The contribution of these diagrams cannot be decomposed into contributions
 we have seen before. The result of each diagram is given by
 \be
T_{66}&=&
 {N \choose  q}^{-3}\sum_{q_1=0}^q \sum_{l_1=0}^q \sum_{m_1=0}^{l_1} \sum_{l_2=0}^q \sum_{m_2=0}^{l_2} (-1)^{q+q_1+m_1+m_2}
    {N-q \choose q-q_1}{q \choose q_1}
    { N-2q+q_1 \choose q-l_1}
\nn \\ &&\times
    { 2q-2q_1 \choose m_1} {q_1 \choose l_1-m_1}
    { N-2q+q_1 \choose q-l_2}{2q-2q_1  \choose m_2}{q_1 \choose l_2-m_2}
\nn\\&=&
    {N \choose  q}^{-3}\sum_{q_1=0}^q \sum_{m_1=0}^{q} \sum_{m_2=0}^{q} (-1)^{q+q_1+m_1+m_2}
    {N-q \choose q-q_1}{q \choose q_1}
    { N-2q+2q_1 \choose q-m_1}
\nn \\ &&\times
    { 2q-2q_1 \choose m_1} 
    { N-2q+2q_1 \choose q-m_2}{2q-2q_1  \choose m_2}.
    \ee
    
    The contribution of diagrams with five intersections to the eighth moment is
    thus given by
    \be
       \frac{M_{8}^{k=5}}{M_2^4} &=& 4 T_{66}.
 \ee

 The most complicated diagram has six intersections, see figure \ref{diak6}.
 The result is given by
 \be
 T_8 &=& {N\choose q}^{-3}\sum_{q_1 =0}^q \sum_{q_2 =0}^q \sum_{m=0}^{q_1} \sum_{n=0}^{2q-2q_1}
 \sum_{k=0}^{m}\sum_{l=0}^{n}\sum_{s=0}^{k}\sum_{t=0}^{l}
(-1)^{q_1+q_2+n+t}
 {N-q \choose q-q_1}{q \choose q_1}
 { 2q-2q_1\choose n}{q_1 \choose m}\nn \\ && \times 
 {m\choose k} {n \choose l}
{N-2q+q_1 \choose q-k-l} 
 {k \choose s}  
 {l \choose t} { q-k-l \choose q_2-s-t} 
 {N-3q+q_1+k+l \choose q-q_2-(m-k+n-l)}.
\label{T8-bn}
 \ee
 Using the convolution property of binomial factors, this can be simplified to
\be
 T_8 &=& {N\choose q}^{-3}\sum_{q_1 =0}^q  \sum_{q_2=0}^q \sum_{n=0}^{2q-2q_1}
\sum_{t=0}^{n}
(-1)^{q_1+q_2+n+t}
 {N-q \choose q-q_1}{q \choose q_1}
 { 2q-2q_1\choose n}{n \choose t}\nn \\ && \times 
{N-2q+2q_1 \choose 2q-q_2-n} 
 {2q-q_2-n \choose q_2-t}
 {2q-2q_2\choose q-q_2}.
\label{T8-simple}
 \ee
 The contribution to the fourth moment with six crossings is thus given by
 \be
 \frac{M_{8}^{E=6}}{M_2^4} &=& T_8. 
\ee
 
 We have checked the above results in several ways. First, when the phase
 factor is eliminated the contribution of each diagram evaluates to one. Second,
the moments agree with numerical
 result of the moments obtained from the eigenvalues of the SYK Hamiltonian
 at finite $N$. Third, for $q=1$ the results agree with the exact analytical
 result in eq. \eref{momq=1}.

 In the next subsection we derive the result for $T_6$ and $T_8$ from
 the general formula \eref{eqn:etaGdCombinatorics}.

\subsection{Calculation of contributions to the moments starting from
  the general formula}
\label{append:genmomex}

\subsubsection{Calculation of $T_6$}
The result of  the diagram (c) of figure \ref{fig1} using the general formula
\eref{eqn:etaGdCombinatorics} is given by
 \be
 T_6 &=& {N\choose q}^{-3}\sum_{a,b,c,p} \frac{(-1)^{a+b+c}N!}{(N-3q+a+b+c+2p)!}
  \nn\\&&\times \frac 1
   {(q-a-b-p)!(q-a-c-p)! (q-b-c-p)!a!b!c!p!}\nn\\
 &=& {N\choose q}^{-3}\sum_{a,b,c,p}(-1)^{a+b+c} {q-a-p \choose b}{q-a -p\choose c}\nn\\
&&\times \frac{N!}{(N-3q+a+b+c+2p)!
(q-b-c-p)!a!p! (q-a-p)!(q-a-p)!}. 
 \ee
 Using $m= b+c$ as new summation variable after summing the first two
 binomials and absorbing $p$ in $a$ 
 this can be rewritten as
\be
 T_6&=& {N\choose q}^{-3}\sum_{a,m,p} {2q-2 a \choose m}
 \frac{(-1)^{a+m}N!}{(N-3q+a+m+p)!
(q-m-p)!(a-p)!p! (q-a)!(q-a)!}\nn \\ 
 &=&{N\choose q}^{-3} \sum_{a,m,p} {2q-2a\choose m}{N-2q+a \choose q-m-p}
 {a \choose p}
 \frac{(-1)^{a+m}N!}{(N-2q+a)!
a! (q-a)!(q-a)!}. \hspace*{0.5cm}
 \ee
 After performing the sum over $p$ we finally obtain
 \be
 T_6 &=& {N\choose q}^{-2}\sum_{a,m}(-1)^{a+m} {2q-2 a \choose m}{N-2q+2a \choose q-m}
 {N-q \choose q-a} {q \choose a},
\ee
which is the result obtained in section \ref{sec:exactMom}.

\subsubsection{Calculation of $T_8$}
In this subsection we derive the result for the diagram $T_8$ starting
from the general result \eref{eqn:etaGdCombinatorics}. We first change from the $d$ representation
to the $c$ representation,
\be
d_{ab} &=& c_{ab} -c_{abc}-c_{abd} +c_{abcd},\qquad d_{ac} = c_{ac} -c_{abc}-c_{acd} +c_{abcd},\nn \\
d_{ad} &=& c_{ad} -c_{abd}-c_{acd} +c_{abcd},\qquad d_{bd} = c_{bd} -c_{bcd}-c_{abd} +c_{abcd},\nn \\
d_{bc} &=& c_{bc} -c_{abc}-c_{bcd} +c_{abcd},\qquad d_{cd} = c_{cd} -c_{bcd}-c_{acd} +c_{abcd},\nn \\
d_{abc}&=& c_{abc}-c_{abcd},\qquad d_{abd}= c_{abd}-c_{abcd},\nn \\
d_{bcd}&=& c_{bcd}-c_{abcd},\qquad d_{acd}= c_{acd}-c_{abcd},\nn \\
d_{abcd}&=& c_{abcd}.
\ee
Then it  can be expressed in terms of binomials as
  \be
  T_8&&={N\choose q}^{-3}\sum(-1)^{c_{ab}+c_{ac}+c_{ad}+c_{bc}+c_{bd}+c_{cd}}
  {N-q\choose q-c_{ab}}{q\choose c_{ab}}
  \nn\\&&\times
  {c_{ab}\choose c_{abc}+c_{abd}-c_{abcd}}
  {c_{abc} +c_{abd} -c_{abcd}\choose c_{abc}}{c_{abc}\choose c_{abcd}}
{c_{ac}-c_{abc}\choose c_{acd}-c_{abcd}}
  {c_{bc}-c_{abc} \choose c_{bcd}-c_{abcd}}
    \nn \\
&&\times 
  {q-c_{ab} \choose c_{ac}+c_{ad}-c_{abc}- c_{abd}-c_{acd}+c_{abcd}}
  {c_{ac}+c_{ad}-c_{abc}-c_{abd}-c_{acd}+c_{abcd} \choose c_{ac}-c_{abc}}
  \nn\\
  &&\times
  {q-c_{ab} \choose  c_{bc}+c_{bd}-c_{abc} -c_{abd}-c_{bcd}+c_{abcd}}
  {c_{bc}+c_{bd}-c_{abc}-c_{abd}-c_{bcd} +c_{abcd}\choose c_{bc}-c_{abc}}
\nn\\ && \times
  {N-3q+c_{ab}+c_{bc}+c_{ac}-c_{abc} \choose q-c_{ad}-c_{bd}-c_{cd}+c_{abd}+c_{acd}+c_{bcd}-c_{abcd}}
  \nn\\ &&\times
  {N-2q+c_{ab} \choose q-c_{ac}-c_{bc}+c_{abc}}
  {q-c_{ac}-c_{bc}+c_{abc} \choose c_{cd}-c_{acd}-c_{bcd}+c_{abcd}},
  \ee
where the sum is over the $c_{a_1\cdots a_k}$.
  Next, we change the summation variables according to
\be
m_1&=& c_{abc}+c_{abd}-c_{abcd},\qquad\\
m_{2a}&=&c_{ac}+c_{ad}-c_{abc}-c_{abd}-c_{acd}+c_{abcd},\qquad
m_{2b}=c_{bc}+c_{bd}-c_{abc} -c_{abd}-c_{bcd}+c_{abcd},\nn\\
s_{2a}&=&c_{acd}-c_{abcd},\qquad
s_{2b}=c_{bcd}-c_{abcd},\qquad
k_{2a}= c_{ac} -c_{abc},\qquad
k_{2b}=c_{bc}-c_{abc},\nn
\ee
and keep the variables $c_{ab}$, $c_{cd}$, $c_{abc}$ and $c_{abcd}$, retaining the same
number of variables.
This results in
\be
  T_8&=&{N\choose q}^{-3}\sum(-1)^{c_{ab}+c_{cd}+m_{2a}+m_{2b}+s_{2a}+s_{2b}  }
  {N-q\choose q-c_{ab}}{q\choose c_{ab}}
  {c_{ab}\choose m_1}
  {m_1\choose c_{abc}}{c_{abc}\choose c_{abcd}}\nn \\
&&\times 
  {q-c_{ab} \choose m_{2a}}
  {m_{2a} \choose k_{2a}}  {k_{2a}\choose s_{2a}}
  {q-c_{ab} \choose m_{2b}}
  {m_{2b} \choose k_{2b}}  {k_{2b}\choose s_{2b}}  {q-k_{2a}-k_{2b}-c_{abc} \choose c_{cd}-s_{2a}-s_{2b}-c_{abcd}}
\nn\\ && \times
  {N-3q+k_{2a}+k_{2b}+c_{ab} +c_{abc} \choose q-m_1-m_{2a}-m_{2b}+k_{2a}+k_{2b}-c_{cd}+c_{abc}}
  {N-2q+c_{ab} \choose q-k_{2a}-k_{2b}-c_{abc}}.
  \ee
  We can now perform the sums over the pairs $\{ s_{2a}, s_{2b}\}$,
  $\{ k_{2a}, k_{2b}\}$ and $\{ m_{2a}, m_{2b}\}$ with constant sum
  for each pair. Introducing the sums
  \be
  s_2&=& s_{2a}+s_{2b},\qquad
  k_2= k_{2a}+k_{2b},\qquad
  m_2= m_{2a}+m_{2b}
  \ee
  as new summation variables we obtain
\be
  T_8&=&{N\choose q}^{-4}\sum(-1)^{c_{ab}+c_{cd}+m_{2}+s_{2}  }
  {N-q\choose q-c_{ab}}{q\choose c_{ab}}
  {c_{ab}\choose m_1}
  {m_1\choose c_{abc}}{c_{abc}\choose c_{abcd}}\nn \\
&&\times 
  {2q-2 c_{ab} \choose m_{2}}
  {m_{2} \choose k_{2}}  {k_{2}\choose s_{2}}
  {q-k_{2}-c_{abc} \choose c_{cd}-s_{2}-c_{abcd}}
  \nn\\
  &&\times
  {N-3q+k_{2}+c_{ab}+ c_{abc} \choose q-m_1-m_{2}+k_2-c_{cd}+c_{abc}}
  {N-2q+c_{ab} \choose q-k_{2}-c_{abc}}.
  \ee
  This is equal to the result \eref{T8-bn} which can be simplified to
  eq. \eref{T8-simple}.

\providecommand{\href}[2]{#2}\begingroup\raggedright\endgroup

\end{document}